\documentclass[usenatbib,onecolumn]{mnras}

\usepackage{newtxtext,newtxmath}

\usepackage[T1]{fontenc}
\usepackage{ae,aecompl}

\usepackage{graphicx}	
\usepackage{amsmath}
\usepackage{amsfonts}
\usepackage{mathtools}
\usepackage{cancel}
\usepackage{subfigure}
\usepackage{graphicx}
\usepackage{float}
\usepackage{color}
\usepackage{bm}
\usepackage{epstopdf}
\usepackage{url}
\usepackage{hyperref}
\usepackage[inline]{enumitem}

\def\prd{Phys.~Rev.~D}

\def\apj{ApJ}
\def\apjl{ApJL}
\def\apjs{ApJS}
\def\aap{A\&A}

\def\apss{Ap\&SS}               

\title{The theory of cosmic-ray scattering on pre-existing MHD modes meets data}

\author[O. Fornieri et al.]
{Ottavio Fornieri,$^{1,2,3}$\thanks{E-mail: ottaviofornieri@yahoo.it}
Daniele Gaggero,$^{3}$\thanks{E-mail: daniele.gaggero@uam.es}
Silvio Sergio Cerri,$^{4}$
Pedro De La Torre Luque,$^{5,6}$
\newauthor
Stefano Gabici,$^{7}$
\\
$^{1}$ Deutsches Elektronen-Synchrotron (DESY), Platanenallee 6, D-15738 Zeuthen, Germany\\
$^{2}$ Department of Physical Sciences, Earth and Environment, University of Siena, Strada Laterina 8, 53100 Siena, Italy \\
$^{3}$ Instituto de F\'isica Te\'orica UAM-CSIC, Campus de Cantoblanco, E-28049 Madrid, Spain \\
$^{4}$ Department of Astrophysical Sciences, Princeton University, Princeton, NJ, United States \\ 
$^{5}$ Istituto Nazionale di Fisica Nucleare, Sezione di Bari, Via Orabona 4, I-70126 Bari, Italy \\ 
$^{6}$ Dipartimento di Fisica ``M. Merlin" dell'Universit\`a e del Politecnico di Bari, Via Amendola 173, I-70126 Bari, Italy \\
$^{7}$ Universit\'e de Paris, CNRS, Astroparticule et Cosmologie, F-75006 Paris, France 
}

\date{Accepted XXX. Received YYY; in original form ZZZ}

\pubyear{2020}

\graphicspath{{./}{figures/}}

\begin{document}
\label{firstpage}
\pagerange{\pageref{firstpage}--\pageref{lastpage}}

\maketitle

\begin{abstract}

We present a comprehensive study about the phenomenological implications of the theory describing Galactic cosmic-ray scattering onto magnetosonic and Alfvénic fluctuations in the $\mathrm{GeV} - \mathrm{PeV}$ domain.
We compute a set of diffusion coefficients from first principles, for different values of the Alfvénic Mach number and other relevant parameters associated to both the Galactic halo and the extended disk, taking into account the different damping mechanisms of turbulent fluctuations acting in these environments.
We confirm that the scattering rate associated to Alfvénic turbulence is highly suppressed if the anisotropy of the cascade is taken into account. On the other hand, we highlight that magnetosonic modes play a dominant role in Galactic confinement of cosmic rays up to $\mathrm{PeV}$ energies.
We implement the diffusion coefficients in the numerical framework of the {\tt DRAGON} code, and simulate the equilibrium spectrum of different primary and secondary cosmic-ray species.
We show that, for reasonable choices of the parameters under consideration, all primary and secondary fluxes at high energy (above a rigidity of $\simeq 200 \, \mathrm{GV}$) are correctly reproduced within our framework, in both normalization and slope. 
\end{abstract}


\section{Introduction}

A diverse set of data, from surveys of the radio and $\gamma$-ray sky to measurements of cosmic-ray chemical composition and anisotropy, has suggested for more than $50$ years the existence of a non-thermal population of cosmic particles confined in our Galaxy with a characteristic timescale of $\sim 10^7$ years (see \citealt{Gabici:2019jvz} and references therein). Such timescale is much larger than the crossing time of the Galactic halo associated to a free-streaming particle, thus suggesting a diffusive transport regime rather than ballistic motion of these cosmic particles.

Resonant interactions between cosmic particles and the Alfv\'enic part of the magneto-hydro-dynamic (MHD) turbulent cascade have been considered as the main origin of this confinement. A perturbative theory able to predict the scattering rate as a function of the particle rigidity (in the limit of small {\it isotropic} perturbations on top of a regular background magnetic field) was developed in the $1960$s~\citep{1966ApJ...146..480J,1966PhFl....9.2377K,1967PhFl...10.2620H,1970ApJ...162.1049H}.
This quasi-linear theory (QLT) of CR scattering on Alfv\'enic turbulence has inspired most phenomenological characterizations of the cosmic-ray sea in terms of a diffusion equation solved by means of numerical or semi-analytical codes like {\tt GALPROP}~\citep{Galprop1,Galprop2,Galprop3}, {\tt DRAGON}~\citep{Evoli:2008dv,Gaggero:2013rya,Evoli2017I,Evoli2017II}, {\tt PICARD}~\citep{Picard1,Picard2} and {\tt USINE}~\citep{Usine}. 
In most of these studies, a simplified, isotropic and homogeneous diffusion equation inspired by QLT is usually implemented, whereby both the normalization and slope of the diffusion coefficient are not determined by first principles, but rather fitted to secondary-to-primary flux ratios (\textit{e.g.}, the Boron-to-Carbon ratio, B/C). 
However, something worth pointing out, is that QLT predicts a highly anisotropic diffusion, that proceeds predominantly in the direction of the magnetic-field lines. Its applications to an isotropic diffusion model is typically justified in terms of large-amplitude turbulent fluctuations of the magnetic field at the scales of their injection~\citep{Strong:2007nh}. However, a rigorous proof that this allows to treat cosmic-ray transport as an effectively isotropic diffusion does not exist to date.

The {\tt DRAGON} package, in particular, has provided some very significant steps forward in this context, \textit{i.e.}, by moving away from the naive zero-order modeling of isotropic, homogeneous diffusion and implementing in some contexts position-dependent diffusion coefficients. Inhomogeneous diffusion has been a key feature of some phenomenogical models aimed at reproducing some well-known anomalies widely debated in the community: for instance, in \citet{Evoli:2012} a variable normalization of the diffusion coefficient provided a solution to the CR {\it gradient problem}; in \citet{Gaggero:2015gradients} a position-dependent scaling of the (isotropic) diffusion coefficient  was invoked to explain a spectral trend inferred by Fermi-LAT data that points towards a progressive hardening of the CR proton spectrum towards the inner Galaxy, with relevant implications for TeV $\gamma$-ray and neutrino astronomy~\citep{Gabici2008APh,Gaggero:2015TeV,Gaggero:2017GC}. In \citet{Cerri:2017} this trend was shown to naturally emerge from different scaling relations for the parallel and perpendicular diffusion coefficients, in the context of a fully anisotropic numerical modeling of cosmic-ray transport in the three-dimensional structure of the Galactic magnetic field~\citep{Jansson:2012rt}. 
In a another series of papers~\citep{Tomassetti:2012ga,Tomassetti:2015mha,tomassetti2017} different scaling relations of the diffusion coefficient in the Galactic disk and in the halo were invoked to explain the hardening in the proton spectrum (first discovered by PAMELA and then characterized by AMS-02 with large accuracy) for a peculiar combination of parameters in the disk and halo. 
These studies have provided a precious insight of the impact of CR trasport properties on both local and non-local observables ($\gamma$-ray, radio, neutrino fluxes) linked to the diffuse population of CR that populate the Galaxy. 

However, as mentioned above, all these studies do not contain a description of cosmic-ray transport that fully captures the microphysics of the interaction between CRs and the magnetized turbulent plasma. A proper implementation of these microphysical processes seems compelling in order to usher in a new era of cosmic-ray modeling, thus providing a proper link between theories and the plethora of increasingly accurate measurements.

From the theoretical point of view, our picture of MHD turbulence and our understanding of CR interactions with the turbulent plasma have dramatically improved during the latest decades with respect to the simple QLT mentioned above. These developments have now led to a more appropriate description of the turbulent cascade in the interstellar medium and its interactions with the diffuse CR {\it sea}.
According to this description, MHD turbulence can be decomposed into a mixed cascade of (incompressible) {Alfv\'enic fluctuations}, {as well as} (compressible) slow and fast magnetosonic {fluctuations}, as theoretically demonstrated and numerically confirmed by simulations~\citep{Cho:2001hf,ChoPhysRevLett.88.245001}.
Regarding the Alfv\'enic component, a reference scenario is the model put forward by Goldreich and Sridhar (hereafter GS95 model, presented in ~\citet{GS1994ApJ...432..612S,GS1995ApJ...438..763G}; see also \cite{2000JPlPh..63..447G,2001ApJ...548..318B}, and \citet{ChoLazarianVishniac2003} for a general review). 
The model stems from the observation that mixing field lines in directions perpendicular to the regular magnetic field on a hydrodynamical-like timescale is easier than bending the lines themselves, because of the magnetic tension. This perpendicular mixing is able to couple wave-like motions that travel along the regular field, obeying a {\it critical balance} condition: $k_\parallel v_A \sim k_\perp u_k$. As the turbulent energy cascades down to smaller and smaller perpendicular scales (larger wavenumbers) with a Kolmogorov-like spectrum, it becomes progressively more difficult for the (weaker and smaller) eddies to bend the field lines and develop {small-scale parallel structures}. Therefore, most of the power is transferred to scales {perpendicular to a mean-magnetic-field direction}, and the model implies a high degree of anisotropy of the Alfv\'enic cascade. These considerations are captured by the scaling relations $E_A(k_\perp) \,\propto\, k_\perp^{-5/3}$ (Kolmogorov-like spectrum in the perpendicular direction), and $k_\| \propto k_\perp^{2/3}$. As shown in \citet{ChoPhysRevLett.88.245001}, the same {anisotropic} scaling relations hold for {a cascade of} slow magnetosonic (or, pseudo-Alfv\'en) {perturbations, while} fast magnetosonic {fluctuations were shown to} feature a {\it isotropic} cascade, with a different scaling of the energy spectrum: $E_M(k) \,\propto\, k^{-3/2}$.  
Moreover, as mentioned in \citet{Ptuskin2006}, all the relevant phases of the interstellar medium (ISM) can be approximated as a low-beta plasma (the plasma $\beta$ parameter is the ratio between the plasma thermal pressure and the magnetic pressure) in almost all cases, although some peculiar regions, such as the interior of super-bubbles for instance, may constitute an exception.
In this regime, fast-magnetosonic modes are less damped than Alfv\'enic fluctuations~\citep[see][and references therein]{BarnesPOF1966}; a result also confirmed by means of (collisionless) kinetic simulations of plasma turbulence showing that, when injecting random magnetic-field perturbations at the MHD scales, magnetosonic-like fluctuations may compete with (and possibly dominate over) the Alfv\'enic cascade as the plasma beta decreases below unity~\citep{CerriAPJL2016,CerriAPJL2017}.

As a consequence of this paradigm, the picture of the microphysics of cosmic-ray pitch-angle scattering may be deeply revised. As shown in  \citet{Chandran2000PhRvL..85.4656C}, the cosmic-ray scattering rates, evaluated for the GS95 highly anisotropic Alfv\'enic spectrum, significantly decrease with respect to the simple assumption of isotropic cascade. On the other hand, the isotropy of the fast-magnetosonic cascade may allow these modes to dominate {CR scattering} for most of the pitch angle range~\citep{PhysRevLett.89.281102}. 

A (weakly) non-linear modification\footnote{Although this approach is sometimes referred to as ``non-linear theory'' of CR scattering, for the sake of clarity we stress that this is not a fully non-linear approach. In fact, the only modification to QLT theory due to non-linear effects consists of a ``turbulent broadening'' of the resonance kernel in the pitch-angle scattering rate \citep[see, e.g.,][]{2009ASSL..362.....S}. 
} of the QLT theory of CR scattering in MHD turbulence has been developed \textit{e.g.} in \citet{1973Ap&SS..25..471V}, and then further reconsidered in \citet{Yan:2007uc} to explicitly address the role of fast-magnetosonic modes; a seminal attempt to implement such treatment in a numerical code, and compare the predictions with a wide set of data, has been recently presented in \citet{Evoli:2013lma}. This theory naturally leads to a set of well-defined predictions for the diffusion tensor, depending on the local ISM properties. 

In fact, the properties of fluctuations' damping associated to different regions of the ISM play a crucial role in the possible suppression of magnetosonic turbulence.
For instance, in an environment such as the magnetized, diffuse halo of our Galaxy, \textit{i.e.}, characterized by very low density ($n_H \sim 10^{-3} \, \mathrm{cm}^{-3}$) and high temperatures ($T \sim 10^6 \, \mathrm{K}$), the mean-free-path associated to Coulomb scattering can be as large as $\sim10^7$ astronomical units~\citep{Yan:2007uc}.
As a result, collisionless (Landau-type) damping is expected to be the dominant process affecting turbulent fluctuations.
On the other hand, in regions where a significant amount of warm ionized hydrogen is present (\textit{i.e.}, the extended Galactic disk, $|z| \lesssim 1 \, \mathrm{kpc}$), the Coulomb collisional
mean free path can be as low as an astronomical unit. 
Hence, viscous damping has to be taken into account, to the point that it may dominate over collisionless damping. 
This in turn affects the relative effectiveness of the pitch-angle scattering rate associated to different MHD modes.
Given this picture, the above-mentioned modified-QLT framework~\citep{1973Ap&SS..25..471V,Yan:2007uc} allows to consistently compute the diffusion coefficients for a wide rigidity range in both environments, and depending on several parameters, including the plasma $\beta$ and the amplitude of the injected turbulent fluctuations.

In this paper, we aim at providing a first phenomenological analysis based on the non-linear extension of the  QLT of cosmic-ray scattering, simultaneously including magnetosonic {and Alfv\'enic} modes. 
By identifying a set of parameters that characterize the ISM properties in the two Galactic regions mentioned above (and thus the relevant damping mechanisms of turbulent fluctuations therein), we compute the associated diffusion coefficients from first principles, following the formalism outlined in \citet{Yan:2004aq,Yan:2007uc}.
We then implement these coefficients in the {\tt DRAGON2} numerical package and test the predictions of the theory against the most recent data provided by the AMS-02 Collaboration. 
In particular, we focus on the fluxes of protons and light nuclei, as well as on the boron-to-carbon flux ratio.

The paper is organized as follows: in Section \ref{sec:scattering_rate_MHD} we describe the general physical setup, leaving the detailed calculations to Appendix~\ref{app:D_mumu_for_MHD}; in Section \ref{sec:diff_coeff_halo_disk} we show how the relevant physical quantities characterizing the diffuse Galactic halo and the extended Galactic disk shape differently the diffusion coefficients within these two environments; in Section \ref{sec:pheno_implications} we show that the computed diffusion coefficients --- implemented in a two-zone model in {\tt DRAGON2} --- can reproduce the primaries' flux spectra, as well as the boron-over-carbon ratio, above $\sim 200 \, \mathrm{GeV}$, for reasonable choices of the physical parameters. Finally, in Section \ref{sec:discussion} and \ref{sec:summary}, we derive the conclusions and discuss some physical implications of these results.

\section{Scattering rate and diffusion coefficient in MHD turbulence}\label{sec:scattering_rate_MHD}

Here we present a summary of the calculation leading to the diffusion coefficient experienced by a cosmic particle with charge $q$ and mass $m$ in a turbulent plasma.
To address the contributions to the scattering efficiency arising from the different MHD cascades (namely, Alfv\'enic and fast/slow magnetosonic), we follow the approach based on the (weakly) non-linear extension --- developed in \citet{1973Ap&SS..25..471V} --- of the original quasi-linear theory of pitch-angle scattering on Alfv\'enic and magnetosonic turbulence~\citep{1966ApJ...146..480J, 1969ApJ...156..445K}.
We refer to Appendix~\ref{app:D_mumu_for_MHD} for the detailed calculations leading to the expressions reported in this Section.

In this formalism, a particle with velocity $\bm{v}$ forming an angle $\theta$ with the background magnetic field $\bm{B}_0$ (\textit{i.e.}, having a pitch angle $\mu\equiv v_\|/|\bm{v}|=\cos{\theta}$) propagating in a turbulent environment whose fluctuations' power spectrum is described by $I$, exhibits a scattering rate in pitch-angle space that can be expressed as~\citep{1969ApJ...156..445K, 1975RvGSP..13..547V}:
\begin{equation}\label{eq:pitch-angle_diffusion_general}
    D_{\mu \mu} = \Omega^2 (1 - \mu^2) \int d^3 \bm{k} \sum^{+\infty}_{n=-\infty} R_n (k_{\parallel} v_{\parallel} - \omega + n \Omega) \left[ \frac{n^2 J_n^2(z)}{z^2}I^{\rm A} (\bm{k}) + \frac{k^2_{\parallel}}{k^2} J'^{2}_{n}(z) I^{\rm M} (\bm{k}) \right],
\end{equation}
where: $\Omega=q B_0 /m \gamma c$ is the particle's Larmor frequency; $\bm{k}$ is the wave-vector of the turbulent fluctuations; $k_{\parallel} \equiv |\bm{k}|\cos{\alpha_{\mathrm{wave}}}$ is its field-aligned component ($\alpha_{\mathrm{wave}}$ being the angle between the wave vector and the direction of the background magnetic field); $\omega=\omega(\bm{k})$ the associated fluctuations' frequency. 

In Equation \eqref{eq:pitch-angle_diffusion_general}, the power spectrum of the fluctuations has been explicitly split into its Alfv\'enic ($I^{\rm A}$) and magnetosonic ($I^{\rm M}$) contribution, since sub-gyro-scale fluctuations belonging to these two components are ``filtered'' differently by particles' gyro-motion: this effect is described by the different coefficients involving the $n$-th order Bessel functions $J_n(z)$ (with $z\equiv k_\perp r_L$, where $r_L= v_\perp/\Omega$ is the particle's Larmor radius\footnote{In the literature, one typically finds $z$ rewritten in terms of the pitch angle $\mu$ and dimensionless rigidity $R=L^{-1}|\bm{v}|/\Omega$ ($L$ being the injection scale of the turbulence), as $z=k_\perp L\, R \sqrt{1-\mu^2}$.}) which (typically) gyro-average out the fluctuations at scales much smaller than the particle gyro-radius ({\em viz.}, $z\gg1$). This, in turn, means also that different fluctuations' components, Alfv\'enic and magnetosonic, differently feed back into particles' motion itself (through the resulting $D_{\mu\mu}$).

Finally, $R_n$ represents a function that ``resonantly'' selects fluctuations whose frequency, in a reference frame that streams along $\bm{B}_0$ with the particle ($\omega'\equiv\omega-k_\|v_\|$), is either zero ($n=0$; Landau-like wave-particle interaction\footnote{{In the case of Alfv\'enic fluctuations, this is the standard Landau damping of Aflv\'en waves, which, however, within this framework does not contribute to the pitch-angle scattering rate to the first order in fluctuations' amplitude, $\delta B_\perp/B_0$. On the other hand, in the case of magnetosonic turbulence, there is a first-order correction to the magnetic-field strength, due to $\delta B_\|$ fluctuations. As a result, there is a non-zero gradient of $|\bm{B}|$ along the field lines, which provides a 
``mirroring force'', $\bm{F}_{\rm mirr}\propto\bm{\nabla}_\||\bm{B}|$, that determines a Landau-like damping, typically referred to as {\em transit-time damping} (TTD). This TTD provides a non-zero contribution to the pitch-angle scattering rate.}}) or matching a multiple (\textit{i.e.}, harmonic) of the particle gyro-frequency ($n=1,2,3,\dots$; gyro-resonant wave-particle interaction).
In the standard QLT treatment of purely Alfv\'enic turbulence, this function is a Dirac $\delta$-function. In the present treatment, instead, we include the effect of turbulent fluctuations on the local magnetic-field strength, \textit{i.e.} the fact that the modulus $|\bm{B}|$ may not be spatially homogeneous. This is clearly dependent on the level of the fluctuations at the injection scale, and is particularly relevant in the presence of magnetosonic (\textit{i.e.}, {\em compressible}) turbulence, whose finite-$\delta B_\|$ fluctuations provide first-order corrections to the magnetic-field strength~\citep{1973Ap&SS..25..471V}. As a result, following \citet{Yan:2007uc}, we adopt a broadened resonance function of the type
\begin{equation}\label{eq:resonant_function_body_text}
    R_n (k_{\parallel} v_{\parallel} - \omega + n \Omega) = \frac{\sqrt{\pi}}{|k_{\parallel}| v_{\perp} M_{\rm A}^{1/2}} e^{- \frac{(k_{\parallel} v \mu - \omega + n\Omega)^2}{k^2_{\parallel} v^2 (1 - \mu^2) M_{\rm A}}},
\end{equation}
where the broadening is determined by the level of the fluctuations through the Alfv\'enic Mach number at the injection scale $L$, $M_{\rm A}\sim(\delta B/B_0)_L$. The resonance function in \eqref{eq:resonant_function_body_text} indeed becomes narrower and narrower as $M_{\rm A}$ decreases (to the point that reduces to a Dirac $\delta$-function, $R_n\to\pi\,\delta(k_\|v_\|-\omega+n\Omega)$, in the limit $M_{\rm A}\to0$).

For the turbulent power spectra $I^{\rm A, M}$, we follow the prescription given in \citet{2002cra..book.....S, PhysRevLett.89.281102} and consider the two-point correlation tensors between the fluctuation components (see Appendix~\ref{app:D_mumu_for_MHD}):
\begin{equation}\label{eq:turbulent_spectra}
    \langle \delta B_i (\bm{k})\,\delta B^*_j (\bm{k'}) \rangle / B_0^2 = \delta^3(\bm{k} - \bm{k'}) \, {\cal M}_{ij}\,,
\end{equation}
where $I^{\rm A, M} = \sum^{3}_{i = 1} {\cal M}_{ii}$ and the spectral scalings are resulting from simulations~\citep{Cho:2001hf,ChoPhysRevLett.88.245001}. 
In particular, we use 

\begin{equation}\label{eq:scaling_alfven_slow_modes}
    I^{\rm A,S} (k_{\parallel}, k_{\perp}) = C_a^{\rm A,S} \, k_{\perp}^{-10/3} \exp(-L^{1/3} k_{\parallel}/k_{\perp}^{2/3})
\end{equation}
for the Alfv\'en and slow modes, consistent with the usual Goldreich-Sridhar (GS95) spectrum~\citep{GS1995ApJ...438..763G}, while for fast modes we use the isotropic spectrum
\begin{equation}
I^{\rm F}(k) = C_a^{\rm F} \, k^{-3/2}. 
\end{equation}

As a final comment on the calculation of $D_{\mu \mu}$, the integral has to be performed up to the \textit{truncation scale} $k_{\mathrm{max}}$ of the turbulence, namely up to the wave-number at which the cascading timescale equals the dissipation scale, as discussed in \citet{Yan:2004aq}. 
In general, the truncation scale for each damping process is a function of the angle $\alpha_{\mathrm{wave}}$, hence the damping mechanism is in general {\it anisotropic}. 
We refer to the next section and to the Appendix for more details on this quantity.

Once all the contributions from the three modes are computed, we can obtain the parallel spatial diffusion coefficient $D$ as a function of the (dimensionless) particle rigidity $R=L^{-1}|\bm{v}|/\Omega$. The expression of $D(R)$ obtained ~\citep{2002cra..book.....S} will be then implemented in {\tt DRAGON} to calculate the propagated particle spectra measured at Earth:
\begin{equation}\label{eq:spatial_diffusion_coefficient}
    D(R) = \frac{1}{4} \int_{0}^{\mu^{*}} d\mu \, \frac{v^2 \, (1 - \mu^2)^2}{D^{\rm M,T}_{\mu \mu} (R) + D^{\rm M,G}_{\mu \mu} (R) + D^{\rm A,G}_{\mu \mu} (R)},
\end{equation}
where $\mu^*$ is the the largest $\mu \in [0, 1]$ for which a particle can be considered as confined by turbulence (\textit{i.e.}, to be in the diffusion regime).
In particular, this means that a CR with rigidity $R$ and pitch angle $\mu$ should undergo a number of scattering ${\cal N}\gg1$ while traveling a distance $L'_{\mathrm{H,D}}$ of the order of a fraction of the scale length $L_{\mathrm{H,D}}$ associated to the Galactic region where it propagates. In this work, we choose such fraction to be $1/5$, defining $L_{\mathrm{H,D}}/5 \equiv L'_{\mathrm{H,D}}$ ($L_{\rm H}$ and $L_{\rm D}$ are the length scales of the diffuse Galactic halo and of the extended Galactic disk, respectively) and assume that they match the typical magnetic-field coherence length in those regions.
In other words, if $\tau_{\rm stream}\sim L'_{\mathrm{H,D}}/v$ is the streaming timescale of a CR across a distance $L'_{\mathrm{H,D}}$, the pitch-angle scattering time of such cosmic particle, $\tau_{\mu\mu}\sim (1-\mu^2)/D_{\mu\mu}$, (\textit{i.e.}, the typical timescale between two consecutive pitch-angle scattering events) must be much shorter than $\tau_{\rm stream}$:
\begin{equation}\label{eq:criterion_diffusive_regime}
    \frac{\tau_{\mu\mu}}{\tau_{\rm stream}}\,\sim\,
    \frac{v}{L'_{\mathrm{H,D}}}\frac{(1-\mu^2)}{D_{\mu\mu}}\, \ll\,1\,.
\end{equation}

Based on this criterion, we observe that $\mu^{*}$ strongly depends on the strength of the turbulence and on the damping parameters, but for $M_{\mathrm{A}} \geq 0.3$ it closely approaches $1$ for all the energies of interest for the present work $(10^{-1} \, \mathrm{GeV} \leq E \leq 10^5 \, \mathrm{GeV})$ in the disk and in the halo, while particles in the disk exit the diffusive regime for $M_{\mathrm{A}} = 0.1$ even at low energy $(E \lesssim 1 \, \mathrm{GeV})$.

\section{Diffusion coefficients in Galactic disk and halo and ISM properties}\label{sec:diff_coeff_halo_disk}

\begin{figure*}
  \includegraphics[width=0.5\textwidth]{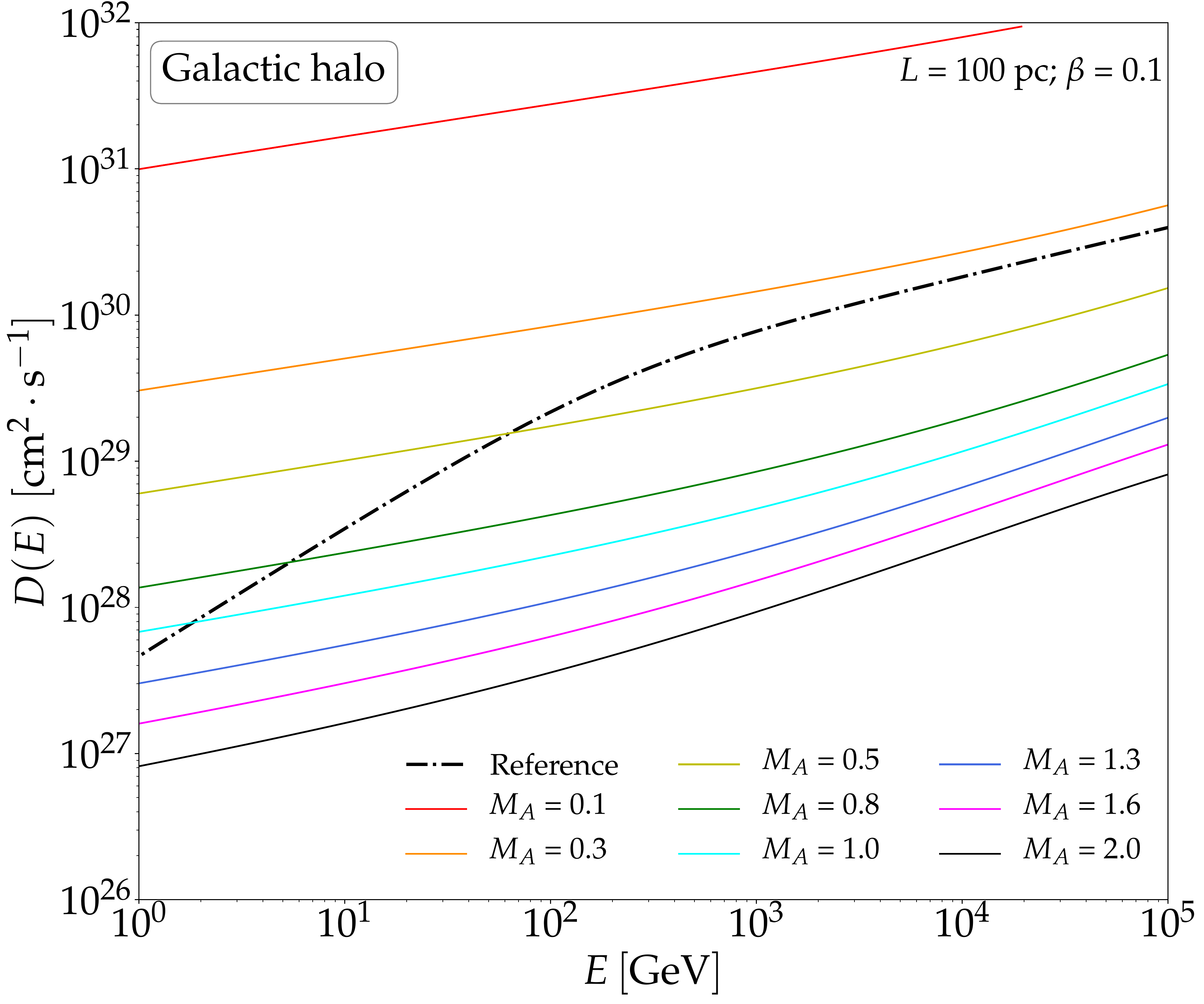}%
  \includegraphics[width=0.5\textwidth]{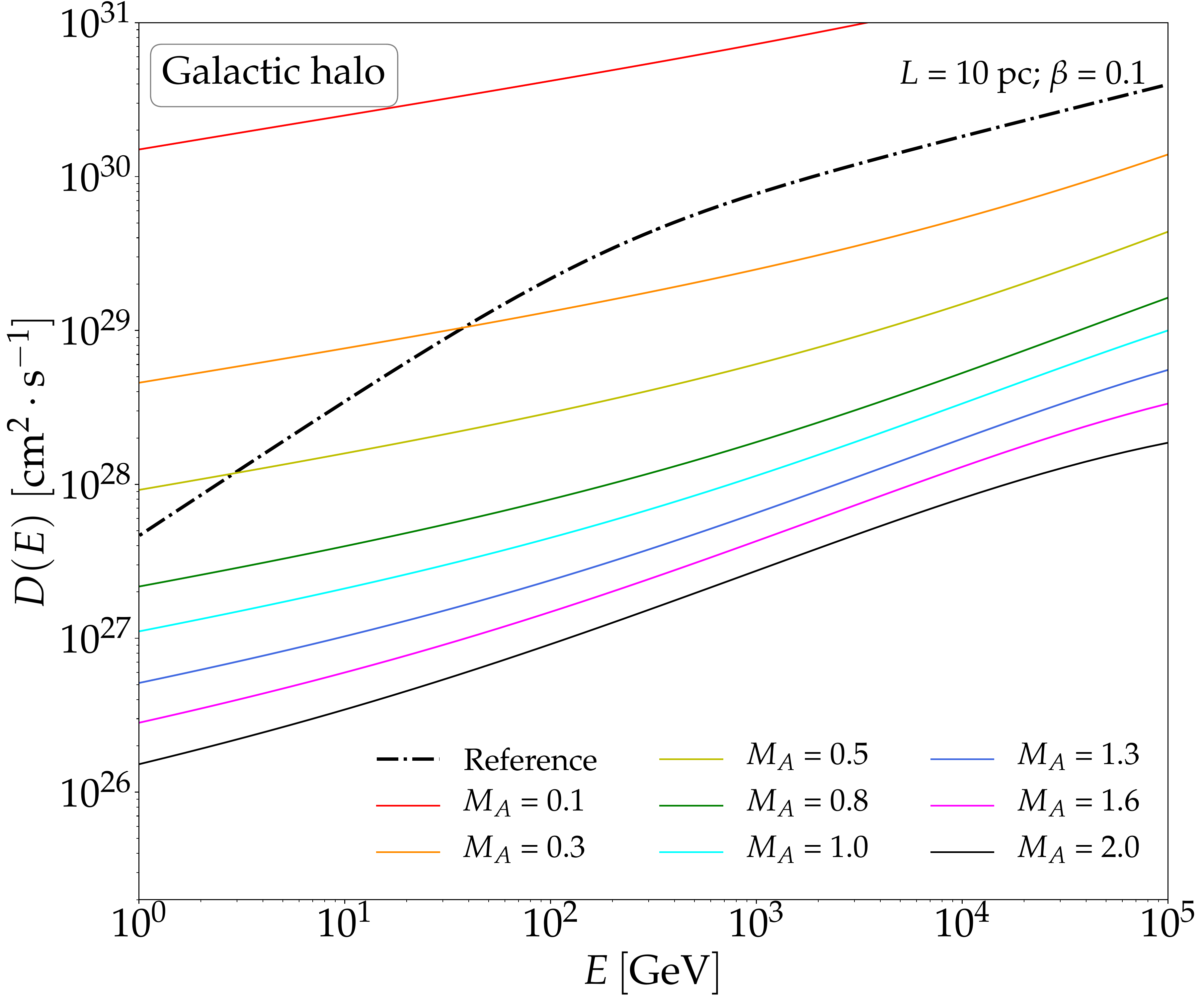}%
\caption{\it {\bf Diffusion coefficients in the halo.} We show the diffusion coefficients associated to the pitch-angle scattering onto MHD (magnetosonic and Alfv\'enic) fluctuations as a function of the rigidity in the halo, given a fixed injection scale $L_{\mathrm{inj}}$ and plasma $\beta$, for several values of $M_{\mathrm{A}}$. Black dashed line: reference diffusion coefficient taken from \citet{Fornieri:2019ddi}, designed to correctly reproduce the AMS-02 data on both primary and secondary species.}
\label{fig:coefficients_halo}
\end{figure*}
\vfill
\begin{figure*}
  \includegraphics[width=0.5\textwidth]{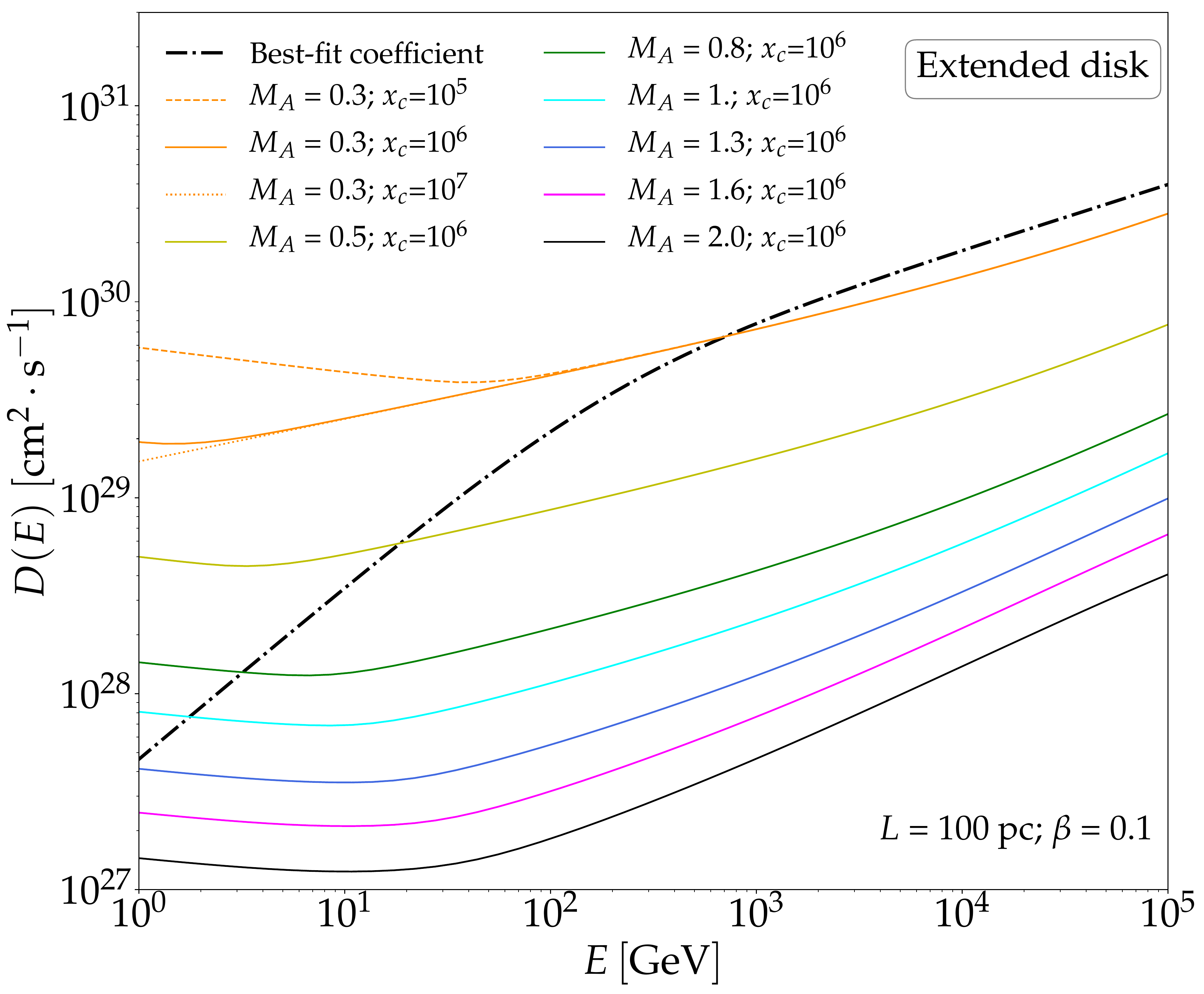}%
  \includegraphics[width=0.5\textwidth]{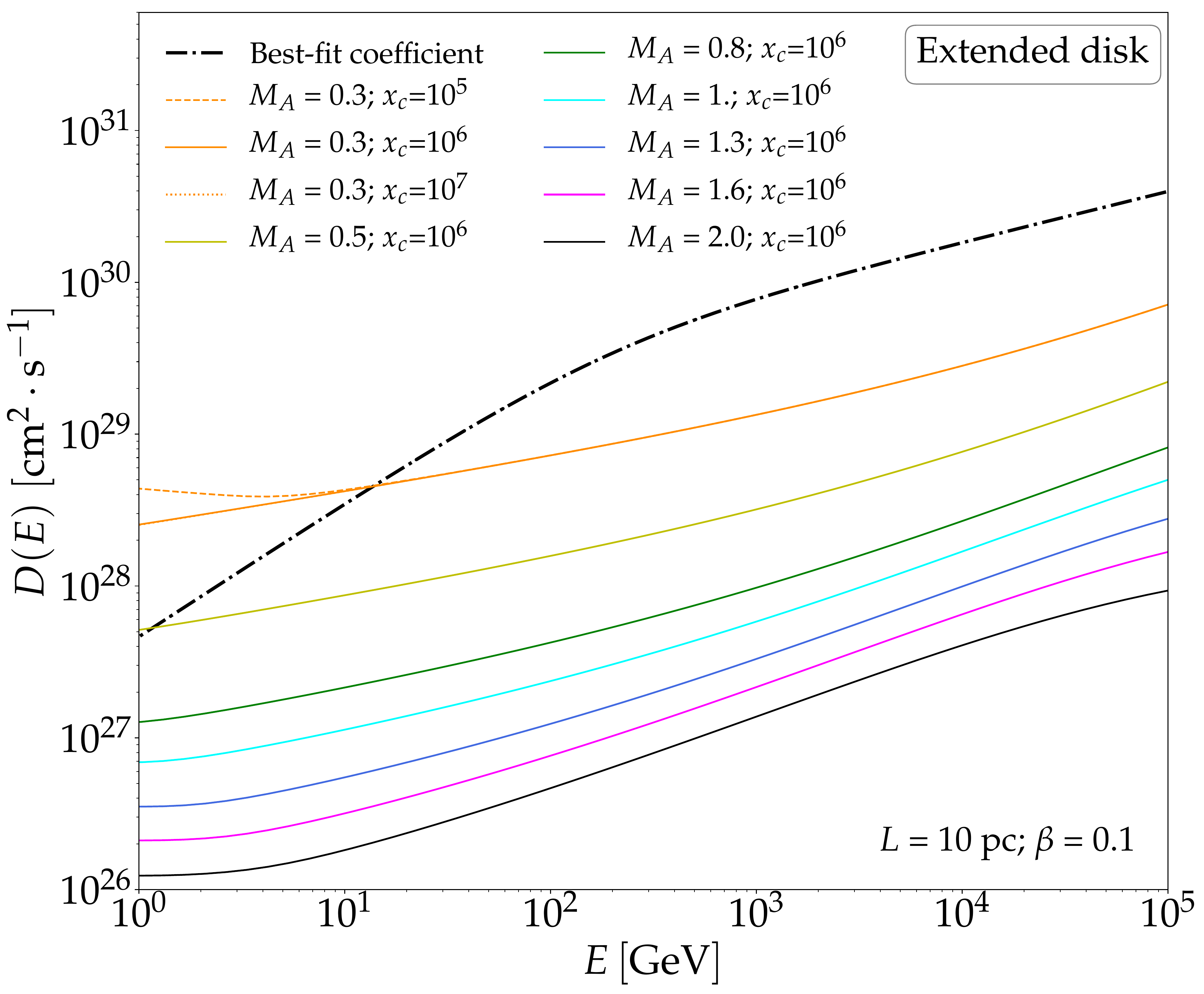}%
\caption{\it {\bf Diffusion coefficients in the extended disk.} As in Figure \ref{fig:coefficients_halo}, we show here the diffusion coefficients associated to the pitch-angle scattering onto MHD (magnetosonic and Alfv\'enic) fluctuations as a function of the rigidity in the ``extended disk'' characterized by effective viscous damping.}
\label{fig:coefficients_disk}
\end{figure*}

In this section, we want to analyze how the diffusion coefficient is shaped by the parameters involved in the calculations.
We take into account two different environments, as sketched in the Introduction: the {\it ``extended disk''}, characterized by the presence of warm ionized hydrogen and a low value of the Coulomb collisional mean free path, and the {\it ``diffuse halo''}, where a low-density plasma characterized by a negligible Coulomb scattering rate is present.

This setup is expected to capture the most relevant phenomenological features of the CR transport problem on Galactic scale. We notice though that the Galactic disk actually exhibits a more complicated structure, and other phases occupy an important fractional volume: namely the {\it hot ionized} phase --- characterized by very low gas density and $T \sim 10^6 \, \mathrm{K}$ --- and the (partially ionized) warm neutral medium. In the latter, ion-neutral collisions may significantly damp both incompressible and compressible modes, hence affect the propagation model discussed here (see for instance the discussion in \cite{Xu2016}). If this complex structure is taken into account, the scattering rate may exhibit strong fluctuations, depending on the ISM phase: This would require another approach to the problem, possibly based on Monte Carlo simulations, that goes beyond the scope of the present work, which is expected to capture instead the main phenomenological consequences of the theory in a large-scale context.

Calculations are carried out using the code in \citet{ottavio_fornieri_2020_4250807}\footnote{\url{https://github.com/ottaviofornieri/Diffusion_MHD_modes}}.

Figure \ref{fig:coefficients_halo} and \ref{fig:coefficients_disk} visualize the diffusion coefficient as a function of the rigidity in the halo and in the disk, respectively, plotted for several values of the Alfv\'enic Mach number $M_{\mathrm{A}}$, given a fixed injection scale $L_{\mathrm{inj}}$ and plasma $\beta$. 
We also show a reference diffusion coefficient taken from  \citet{Fornieri:2019ddi}, designed to correctly reproduce the AMS-02 data on both primary and secondary species.

First of all, we notice that (i) the high-rigidity slope {\it predicted} by the theory (and fixed by the scaling of the fast magnetosonic cascade) is perfectly compatible with the high-rigidity slope of the reference diffusion coefficient fitted on CR data, and (ii) the theory predicts a clear departure from a simple power law for all values of the relevant parameters; however, this departure does not describe the low-energy downturn of the reference coefficient, that reflects the behaviour of AMS-02 data. Hence, we may argue that the theory may provide a correct description of CR confinement above $\simeq 200 $ GV, while an accurate low-energy treatment may require additional theoretical arguments. This argument will be further developed in the next Section. 
The normalization span several orders of magnitude; it is important to notice that it is mainly governed by the value of $M_{\mathrm{A}}$, and that reasonable values of this parameter are associated to the correct normalization.

We will now elaborate more on this aspect and discuss the following key points: (i) the behaviour with respect to the Alfv\'enic Mach number, that reflects the strength of the injected turbulence, (ii) the features associated to the different damping mechanisms involved in the process, and (iii) the role of the Alfv\'en modes. 
The effect of variations on the plasma $\beta$ parameter and the injection scale, $L_{\mathrm{inj}}$ will be also briefly addressed. 

\vspace{0.3cm}
\hspace{-0.6cm}\textbf{\textit{$\bm{D(E)}$ variation with $\bm{M_{\mathrm{A}}}$}}. Both figures clearly show that $D(E)$ is a decreasing function of the Alfv\'enic Mach number. 
This is due to the fact that an increased level of turbulence results in a more effective scattering rate of cosmic particles. 
In fact, by definition $M_{\rm A} \equiv \delta u/v_{\rm A}$: therefore, larger values of $M_{\mathrm{A}}$ characterize larger-amplitude turbulent fluctuations that enhance the pitch-angle scattering rate, $D_{\mu\mu}$. 
As a result, CRs are more efficiently confined at high-$M_{\rm A}$, which results in a lower spatial diffusion coefficient, $D(E)$.

\vspace{0.3cm}
\hspace{-0.6cm}\textbf{\textit{Effect of damping}}. The most relevant difference between the behaviour of $D(E)$ in the halo (Figure \ref{fig:coefficients_halo}) and in the extended disk (Figure \ref{fig:coefficients_disk}) is the minimum in the low-energy domain ($\rho_{\mathrm{min}} \sim  50-100 \, \mathrm{GV}$) in the latter case. 
\begin{figure}
  \centering
  \includegraphics[width=0.45\textwidth]{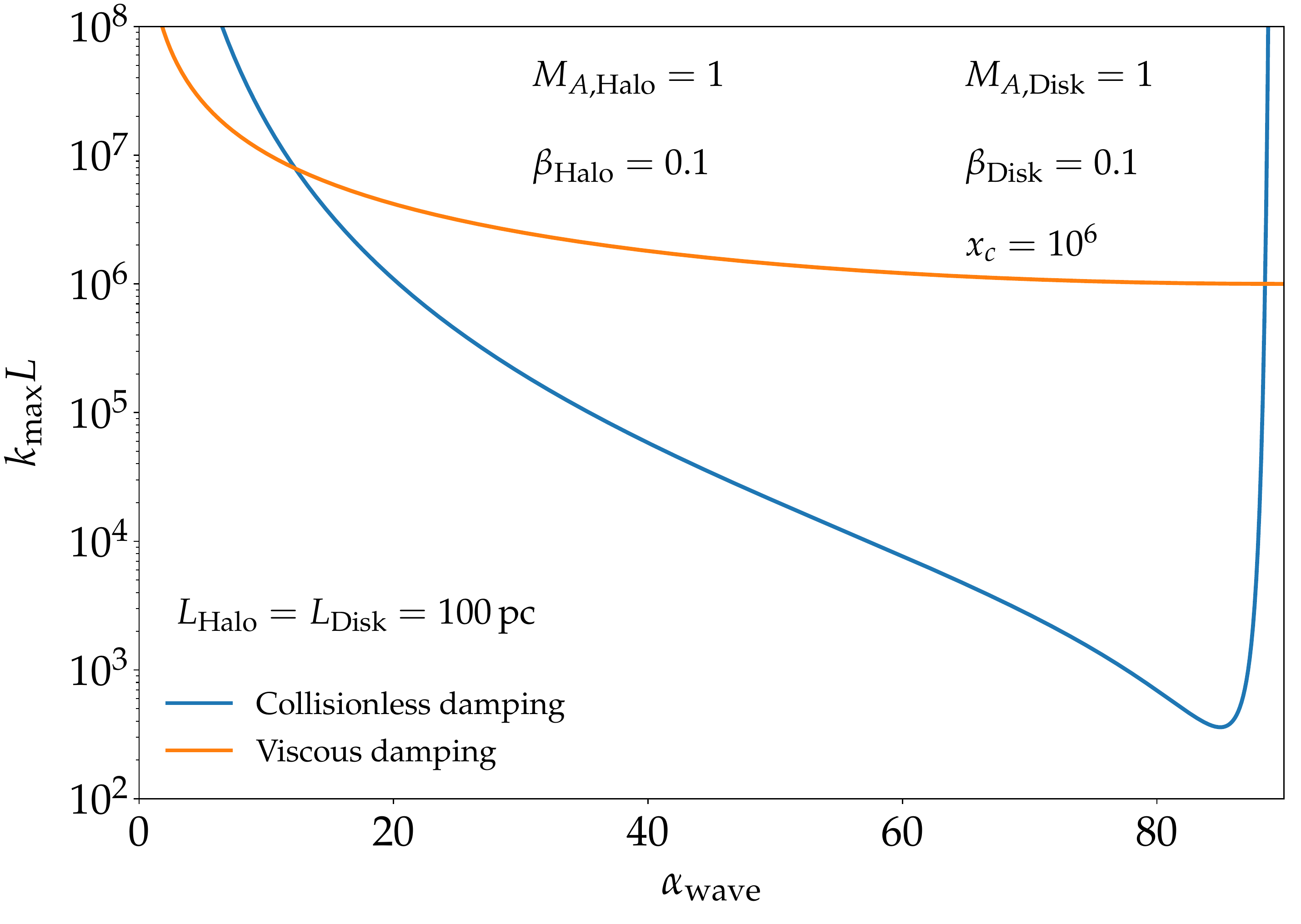}
\caption{\it {\bf Effect of Damping.} We show the truncation scale $k_{\mathrm{max}}$ of the scattering-rate integral as a function of the pitch angle of the turbulent wave with respect to $\bm{B}_0$ for the different damping processes considered in this work. Viscous damping is effective in the extended disk only. The values of the corresponding physical quantities are shown in the plot.}
  \label{fig:truncation_scale}
\end{figure}

\begin{figure*}
  \centering
  \includegraphics[width=0.74\textwidth]{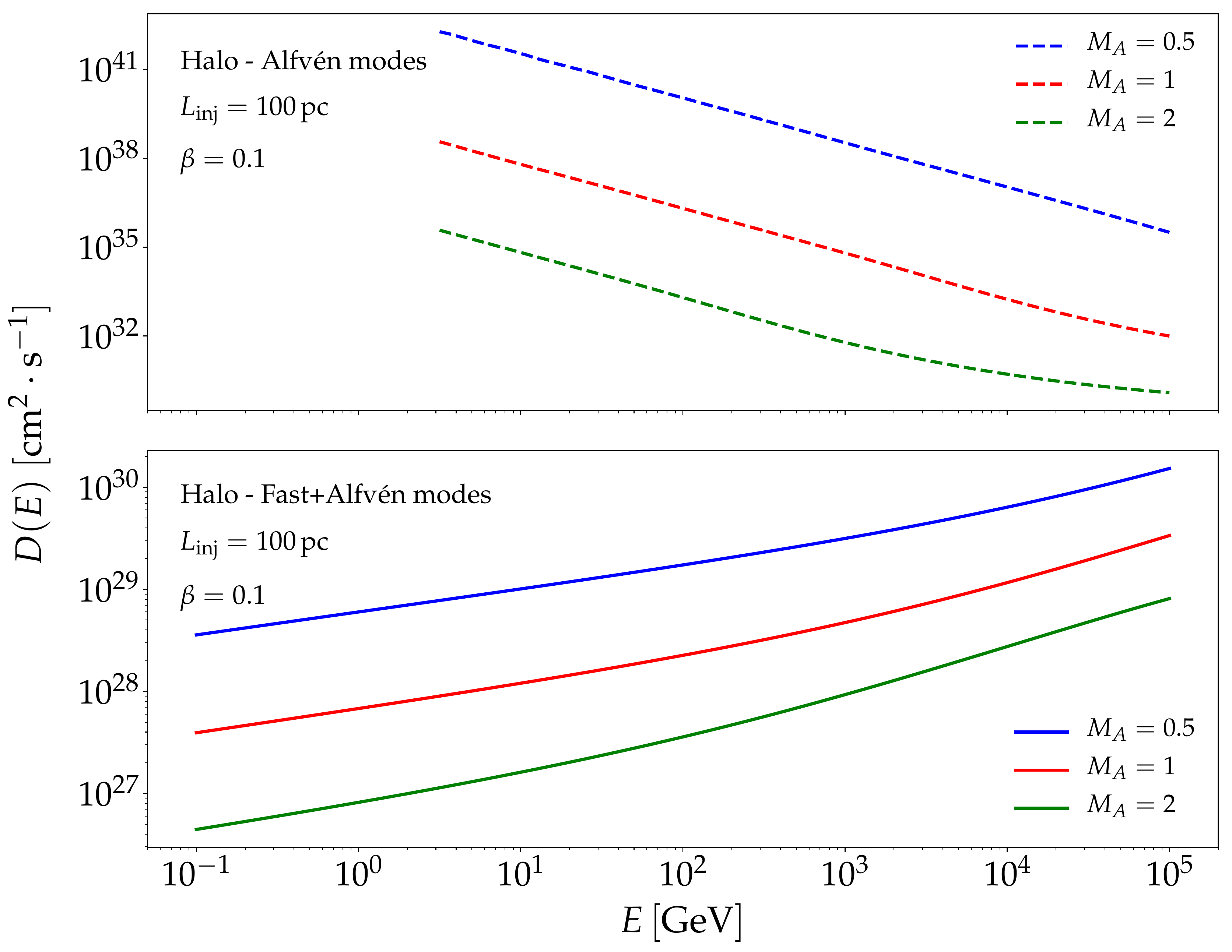}
\caption{\it {\bf Ineffectiveness of Alfvénic confinement.} We show the total diffusion coefficient  with fast magnetosonic modes included in the calculation (lower panel) compared to the case in which only Alfvén fluctuations are taken into account (upper panel). Alfvénic turbulence is not efficient in confining Galactic CRs, due to the anisotropy of the cascade (see also \citet{Chandran2000PhRvL..85.4656C,Yan:2004aq}).} 
\label{fig:coefficients_halo_AllMHD_vs_alfvenOnly} 
\end{figure*}

This feature can be explained following this train of thoughts.
\begin{itemize}
    \item As mentioned in the previous Section, the expression for $D_{\mu \mu}$ involves an integral in the wave vector space $d^3 \bm{k}$ up to a truncation scale $k_{\mathrm{max}}$. This integral is dominated by the contributions associated to waves with small angle $\alpha_{\mathrm{wave}}$ with respect to the direction of the regular magnetic field~\cite[see][]{Yan:2004aq}.
    \item The truncation scale as a function of $\alpha_{\mathrm{wave}}$ associated  to the collisionless damping (present in both the extended disk and in the halo), and to the viscous damping (present in the extended disk only) is shown in Figure \ref{fig:truncation_scale}. In the critical region associated to small angles, the truncation scale associated to collisionless damping is much larger than the one associated to viscous damping.
    \item As a consequence, {\it in the extended disk} environment, the truncation of the scattering-rate integral over $d^3 \bm{k}$ at relatively small wavenumbers ($k_{\mathrm{max}} L \sim 10^7$) implies a lower value of $D_{\mu \mu}$ for CRs at the low energies, the ones that would resonate with comparable (or larger) wavenumbers. This is reflected in the low-rigidity upturn of the {\it spatial} diffusion coefficient shown in Figure \ref{fig:coefficients_disk}. It can also be easily understood that the position of this upturn shifts in energy depending on the intersection point of the two truncation-scale curves in Figure~\ref{fig:truncation_scale}.
\end{itemize}
\vspace{0.3cm}
\hspace{-0.6cm}\textbf{\textit{Role of the Alfvén modes in the confinement process}}. Here, we want to comment on the importance of the fast magnetosonic modes in confining charged CRs. 
In Figure \ref{fig:coefficients_halo_AllMHD_vs_alfvenOnly} we show the diffusion coefficient when fast modes are included (lower panel) compared to the case where only Alfvén modes enter the calculation (upper panel).

Studying the case with no fast modes, two features are immediately visible: 
\begin{itemize}
\item The normalization of $D(E)$ spans from just a few up to $\sim 15$ orders of magnitude more than the case where fast modes are included. Based on the abundances and average lifetimes of unstable elements, the average residence time of CRs in the Galaxy is found to be $\tau_{\mathrm{esc}} \simeq 15 \, \mathrm{Myr}$ in the $\mathrm{GeV}$ domain~\citep{YANASAK2001727}. This implies that, in order to be confined in a halo of a few $\mathrm{kpc}$, CRs should experience a diffusion coefficient that can be at most $\langle D \rangle = \frac{L_{\rm H}^2}{2 \tau_{\mathrm{esc}}} \sim 10^{30} \, \mathrm{cm^2/s}$. 
Therefore, if only Alfvén modes were responsible for confinement, the current data on secondary and unstable species would not be reproduced. Moreover,  the scattering rate would be so low that the diffusion approximation would not be valid anymore, and the CR ``sea'' would be highly suppressed due to ballistic escape from the Galaxy. 
\item {The behaviour of the diffusion coefficient with rigidity shows a declining trend in the pure Alfvénic case, while the total coefficient {\it increases} with rigidity.}
\end{itemize}

Both features derive from the anisotropic behavior of the alfvénic cascade. Indeed, as shown in Equation \eqref{eq:scaling_alfven_slow_modes}, Alfvén modes cascade anisotropically, evolving on the isosurfaces identified by $k_{\parallel} \propto k_{\perp}^{2/3}$~\citep{GS1995ApJ...438..763G}. This relation implies that turbulent eddies are spatially elongated along $\bm{B}$, or, equivalently, that in the momentum space they are elongated in the perpendicular direction. So the majority of the power goes into a $k_{\perp}$ cascade. This leaves very little power (\textit{i.e.} scattering efficiency) to the cascade in parallel wave numbers $k_{\parallel}$ that, according to the resonance function \eqref{eq:resonant_function_body_text}, is the component involved in the wave-particle interaction. Since $k_{\parallel} \sim \ell_{\parallel}^{-1}$, particles with small rigidity and small Larmor radius --- interacting with large $k_{\parallel}$ --- get weakly confined, while high-energy CRs scatter more efficiently. As a result, the spatial diffusion coefficient $D(E)$ is shaped as a {\it decreasing} function of the energy, if only the Alfvénic component is taken into account.

Therefore, an efficient wave-particle scattering with Alfvén modes can occur only at high energies, that resonate with scales that are not too far from the injection scale, where the anisotropic nature of the cascade has not become significant yet.
We can have an estimate of this scale, by computing for instance how many $k_{\perp}$-orders the cascade has to evolve in order to change $k_{\parallel}$ of one order of magnitude. Indeed, as already said $k_{\parallel} \propto k_{\perp}^{2/3}$, which means that the spectral anisotropy of the fluctuations increases as
\begin{equation*}
    \frac{k_{\parallel}}{k_\perp} \sim \left(\frac{k_{\perp}}{k_{\rm inj}}\right)^{-1/3}\,,
\end{equation*}
where we denoted with $k_{\rm inj}$ the (isotropic) wavenumber associated to the injection scale, $L_{\rm inj}$.

As a safe estimate, we can consider the cascade anisotropy to be really important when there is roughly an order of magnitude between the parallel and perpendicular wave numbers corresponding to the same level of turbulent energy, \textit{i.e.},  $k_\|/k_\perp\sim1/10$.
According to the above relation, this level of cascade anisotropy is reached at $k_\perp/k_{\rm inj}\sim10^3$. If we now consider an injection scale $L_{\mathrm{inj}} \sim 100\, \mathrm{pc}$, this will happen at $\ell_{\mathrm{an}} \sim 10^{-3} L_{\mathrm{inj}} \simeq 0.1 \, \mathrm{pc}$. The Larmor radius of a charged CR is $r_L = 3.37 \cdot 10^{12} \, \mathrm{cm} \left( p/\mathrm{GeV} \right) \simeq 1.08 \cdot 10^{-6} \, \mathrm{pc} \left( p/\mathrm{GeV} \right)$. Therefore, a $\ell_{\mathrm{an}} \sim 0.1 \, \mathrm{pc}$ scale roughly corresponds to the Larmor radius of particles belonging to energies $\sim 10^5 \, \mathrm{GeV} \sim 100 \, \mathrm{TeV}$. (Note, however, that considering the anisotropy to be important at $k_\|/k_\perp\sim1/10$ is quite arbitrary, and one may push the above constraint to even larger energies by considering, \textit{e.g.}, $k_\|/k_\perp\sim1/3$ to be already relevant -- this would correspond to CR energies of $\sim3\,\mathrm{PeV}$.)
As a consequence, we would not observe any contribution to $D(E)$ at energy scales that are currently of interest. If, on the other hand, turbulence is injected at smaller scales --- say $L_{\mathrm{inj}} = 10 \, \mathrm{pc}$ for instance --- the same effect comes into play at smaller scales, which therefore contains non-negligible scattering power even at CR energies that are low enough to be experimentally explored ($E \sim 10^4 - 10^5 \, \mathrm{GeV}$). This is indeed visible in the change of slope in $D(E)$ for the larger Mach numbers of Figures \ref{fig:coefficients_halo} and \ref{fig:coefficients_disk} (in the right panels, corresponding to $L = 10 \, \mathrm{pc}$).

This is of course only a rough estimation, since it depends on the strength of the injection --- related to the value of $M_{\mathrm{A}}$ --- and holds as soon as the critical balance is reached and the cascade follows the GS95 spectrum. This would happen at the scale $\ell_{\mathrm{tr}} \sim L_{\rm inj} M_{\mathrm{A}}^2$ or at $\ell_{\rm A}\sim L_{\rm inj}/M_{\rm A}^3$ for sub-Aflv\'enic ($M_{\rm A}<1$) or super-Alfv\'enic ($M_{\rm A}>1$) injection, respectively~\citep{Lazarian:2020cms}, \textit{i.e.}, at scales smaller than $L_{\rm inj}$ if $M_{\rm A}\neq1$. So it is a reasonable estimation for $M_{\mathrm{A}}\approx1$ and this is why there is no imprint of a change of slope in the blue and red dashed lines in the upper panel of Figure \ref{fig:coefficients_halo_AllMHD_vs_alfvenOnly}. By increasing the strength of the injection (\textit{i.e.}, increasing $M_{\rm A}$), anisotropy starts to play a role at lower and lower energies, as exhibited in the green dashed line of the figure. However, there are indications that typical values of the Alfv\'enic Mach number in the ISM do not significantly exceed $M_{\mathrm{A}} \approx 2$~\citep{2011ApJ...736...60T}.

In conclusion, for the injection scale $L_{\mathrm{inj}}$ and Alfv\'enic Mach number we are considering throughout this work, anisotropy of the Alfvén cascade always plays a key role and therefore cannot efficiently confine cosmic rays, while the fast magnetosonic cascade is able to induce a very efficient pitch-angle scattering rate.

\vspace{0.3cm}
Another important parameter to be monitored is the size of the extended disk and Galactic halo. The Galactic halo size determines the volume where cosmic rays propagate, thus influencing the normalization of the diffusion coefficient. Variations on these parameters are important when computing the total diffusion coefficient at a given position in the Galaxy. In general, what is expected to matter is the relation between their sizes. While the halo half-size could be constrained to be between $3-12 \, \mathrm{kpc}$~\citep{di2013cosmic, zaharijas2012fermi, CarmeloBeB}, the extended disk half-size could vary from $0.5$ to $2 \, \mathrm{kpc}$~\citep{Tomassetti_MC}. Along the paper, we will refer to the size of these extended zones as their half-size, \textit{i.e.} a halo size of $L_{\rm H}$ means that it extends from $-L_{\rm H}$ to $+L_{\rm H}$ in the vertical (perpendicular to the Galactic plane) coordinate.

Finally, variations of the plasma beta parameter lead to more efficient confinement of charged particles (\textit{i.e.}, a smaller diffusion coefficient) as $\beta$ decreases (due to the fact that the fast-magnetosonic mode becomes progressively more important in the confinement\footnote{This is because fast-magnetosonic modes become less and less damped at lower beta~\citep[cf., \textit{e.g.},][]{BarnesPOF1966,CerriAPJL2016}. This feature can be appreciated through the behaviour of their {\em collisionless} truncation scale with $\beta$ (see Appendix~\ref{app:D_mumu_for_MHD}).}). This will be shown in the next Section.

To summarize, these calculations allow us to examine how plausible plasma properties characterizing the different Galactic zones can lead to different values of the diffusion coefficient and, therefore, to different spectra of Galactic cosmic-ray fluxes. Different combinations of the plasma parameters in the extended disk and Galactic halo will be explored in the next Section in comparison to experimental data.

\section{Phenomenological implications of the theory}\label{sec:pheno_implications}

In this Section we compare the propagated CR spectrum, obtained adopting the diffusion coefficients discussed above, with the most relevant CR observables.

We implement the coefficients in the {\tt DRAGON2} code~\citep{Evoli:2008dv,Evoli2017I,Evoli2017II}, and solve the diffusion-loss equation that characterize the propagation of high-energy charged particles in the Galaxy~\citep{1964ocr..book.....G,Ginzburg:1990sk}:

\begin{equation}\label{eq:prop}
\begin{split}
- \bm{\nabla} \cdot ( D \bm{\nabla} N_i \,+\, \bm{v}_w N_i) + \frac{\partial}{\partial p} \left[ p^2 D_{pp} \, \frac{\partial}{\partial p} \left( \frac{N_i}{p^2} \right) \right] - \frac{\partial}{\partial p} \left[ \dot{p} N_i - \frac{p}{3} \left(\bm{\nabla} \cdot \bm{v}_w \right) N_i \right] = \\
Q + \sum_{i<j} \left( c \beta n_{\rm gas} \, \sigma_{j \rightarrow i} + \frac{1}{\gamma \tau_{j \rightarrow i}} \right) N_j - \left( c \beta n_{\rm gas} \, \sigma_i + \frac{1}{\gamma \tau_i} \right) N_i
\end{split}
\end{equation}
for the most relevant hadronic species characterized by the distribution function $N_i$, and for different choices of the free plasma parameters identified in the previous Sections. In the above equation, $D$ is the spatial diffusion coefficient; $D_{pp}$ the diffusion coefficient in momentum space, associated to reacceleration; $\bm{v}_w$ the velocity associated to the advection; $(\sigma_{j \rightarrow i}, \sigma_i)$ the spallation cross sections associated, respectively, to the creation of the species $i$ from parent nucleus $j$, and to the destruction of the species $i$; $(\tau_{j \rightarrow i}, \tau_i)$ the decay times for, respectively, the unstable species $j$, creating $i$, and for $i$, creating smaller nuclei. For a detailed discussion on each term, we remind to the technical papers cited above.

The aim of this work is to identify and comment on relevant trends, and discuss whether the current data have some constraining power on the theory; we postpone a complete and systematic exploration of the parameter space to a forthcoming publication.

We adopt the following setup:

\begin{itemize}
    \item {\bf Source term}: We consider a continuous source distribution in two dimensions --- assuming cylindrical symmetry --- taken from \citet{Ferriere:2001rg}. This distribution accounts for the spatial distribution of both type Ia (traced by old disk stars), and type II (traced by pulsars) supernovae. The injection spectrum for each species is modeled as a simple power law in rigidity.
    \item {\bf Gas distribution}: We implement a smooth, cylindrically symmetric gas distribution, taken from \citet{1976ApJ...208..346G,Bronfman1988} and implemented in the public version of both {\tt GALPROP}~\citep{Galprop1,Galprop2,Galprop3} and {\tt DRAGON}.
    \item {\bf Spallation cross sections}: We use the cross-section network presented in \citet{Evoli2017II,2019PhRvD..99j3023E} and implemented in the {\tt DRAGON2} version available online\footnote{\url{https://github.com/cosmicrays}}. This secondary production model is especially designed to compute the light secondary fluxes above $1 \, \mathrm{GeV/n}$, and is based on a state-of-the-art fitting of a semi-empirical formalism to a large sample of measurements in the energy range $100 \, \mathrm{MeV/n}$ to $100 \, \mathrm{GeV/n}$, taking into account the contribution of the most relevant decaying isotopes. 
\end{itemize}

\begin{figure*}
  \includegraphics[width=0.5\textwidth]{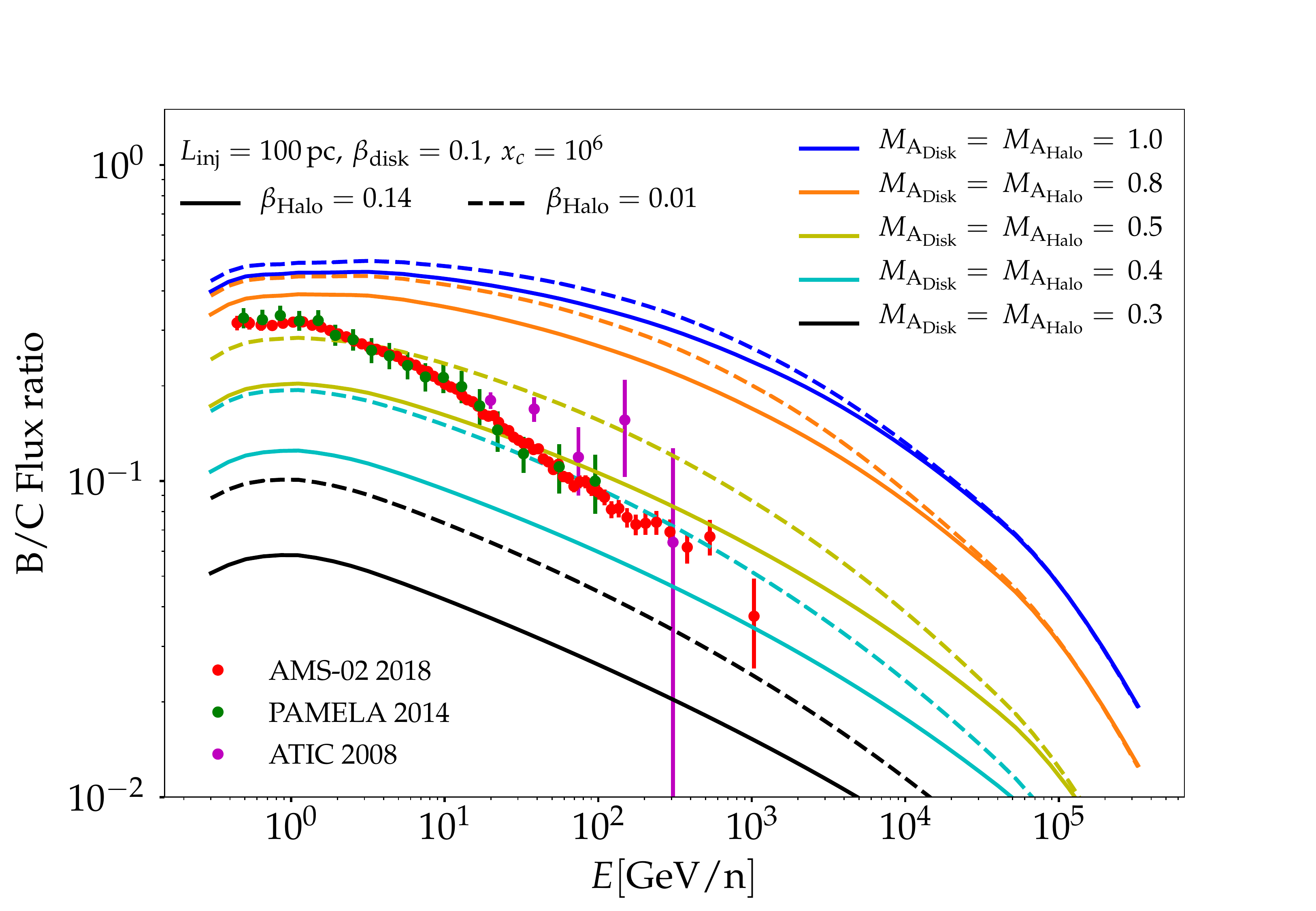}%
  \includegraphics[width=0.5\textwidth]{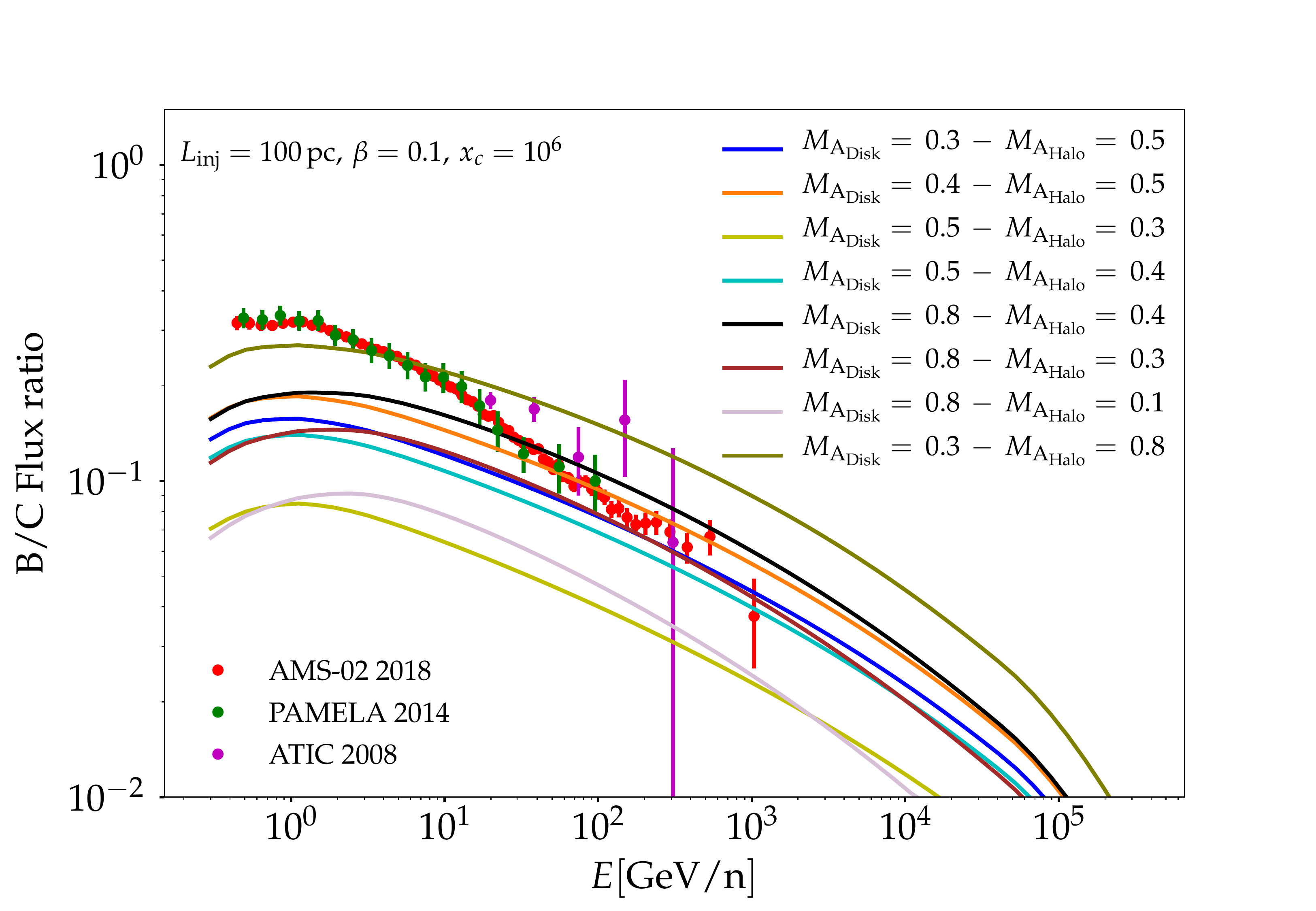}%
\caption{\it {\bf B/C ratio as a function of $M_{\mathrm{A}}$.} We plot the theoretical prediction (obtained with an updated version of the {\tt DRAGON} code) for the B/C within {\it simple setup} characterized by the same value of $M_{\mathrm{A}}$ in both the extended disk and the halo up to TeV energy (left panel), and a more general setup where extended disk and halo exhibit different values of this parameter (right panel). 
We show the most recent data in the energy range of interest from AMS-02, PAMELA and ATIC experiments.}
\label{fig:BC_MAbeta_final}
\end{figure*}
\begin{figure}
  \centering
  \includegraphics[width=0.5\textwidth]{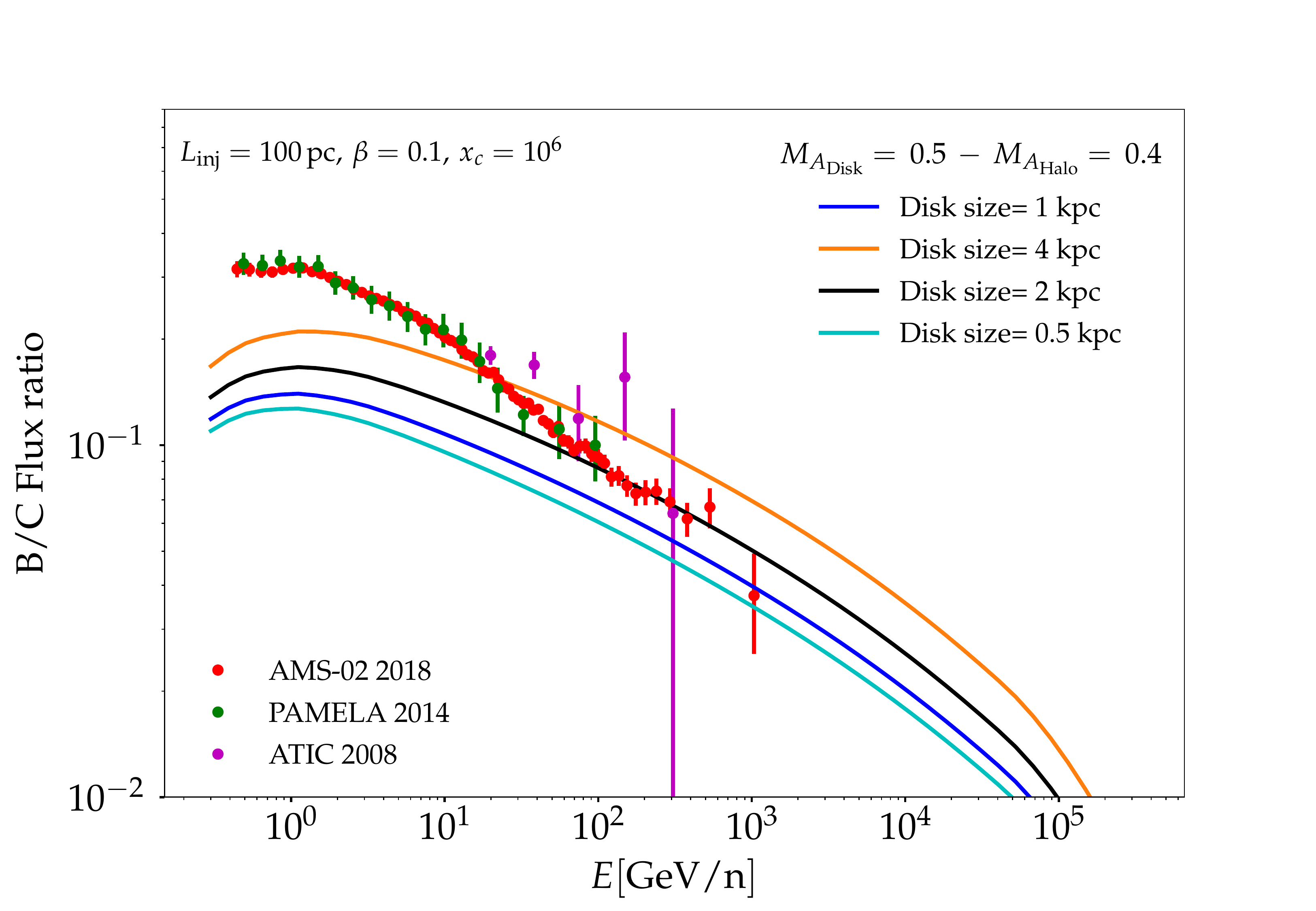}
\caption{\it {\bf B/C ratio as a function of $M_{\mathrm{A}}$.} We plot the theoretical prediction for the B/C for different values of the extended disk vertical size.} 
  \label{fig:BC_DS} 
\end{figure}

A key observable in the context of CR phenomenology is the Boron-over-Carbon (B/C) flux ratio. In fact, Boron is entirely secondary and is mostly produced in spallation reactions involving
heavier, and mostly primary, species (including Carbon): therefore, the ratio between those two nuclei fluxes has been widely used over the latest years to constrain the {\it grammage} accumulated by CRs while residing in the Galactic disk, and ultimately the features of the diffusion coefficient.

Given these considerations, we start our analysis by focusing on this observable, recently measured with high accuracy all the way up to the TeV scale by the AMS-02 Collaboration~\citep{PhysRevLett.117.231102}.
In particular, we pay attention to the dependence of the computed B/C flux ratio on the Alfv\'enic Mach number parameter of pre-existing MHD turbulence, $M_{\mathrm{A}}$, which was shown to play a key role in the overall normalization of the transport coefficients.
We scan over this parameter, and find that larger values of $M_{\rm A}$ are likely to be associated with a significant over-production of Boron, especially at high energies. This is due to the high efficiency of the confinement mechanism that characterize scenarios featuring turbulence with large Alfv\'enic Mach numbers.

\begin{itemize}
\item In a {\it simple setup} characterized by the same value of $M_{\mathrm{A}}$ in both the extended disk and the halo, we find that values of order $M_{\rm A}\sim 0.4$ for the effective Alfv\'enic Mach number are compatible with current data in the high-energy range (see Figure \ref{fig:BC_MAbeta_final}, {\it left panel}). 
We emphasize that this result is achieved with no {\it ad hoc} retuning on the data, and naturally stems from the theoretical expression of the diffusion coefficient computed in detail in this work. 
\item In a {\it more general setup} where extended disk and halo exhibit different values of this parameter, a diverse range of combinations is allowed by the data (see Figure \ref{fig:BC_MAbeta_final}, {\it right panel}). 

We also show for illustrative purposes in Fig. \ref{fig:BC_DS} the impact of the extended disk size on the same observable, keeping the Alfvénic Mach number in the extended disk and halo fixed to one of the combinations allowed by data.

We remark again that in all cases the high-energy slope is correctly reproduced, while the low-energy domain suggests an extra grammage possibly associated to a different confinement mechanism (not captured by the theory presented here) that starts to dominate below $\sim 200\, {\rm GeV}$. This point will be further discussed below.
\end{itemize}

We now widen our perspective and consider a variety of secondary and primary species.

The AMS-02 Collaboration has recently measured the spectra of several CR light nuclei fluxes and ratios~\citep{PhysRevLett.114.171103,PhysRevLett.117.231102,PhysRevLett.121.051103}. These data provided major improvement in the precision and dynamical range, and have revealed relevant features. 
The most relevant is a progressive hardening in primary species, with the spectral index varying from $\simeq 2.8$ in the $50$ - $100 \, \mathrm{GV}$ rigidity range to a significantly harder value around $\simeq 2.7$ above $200 \, \mathrm{GV}$. 
Regarding the primary species, we emphasize that the slopes of the primary species depend on both the rigidity scaling of the diffusion coefficients, and on the slope that is injected in the interstellar medium as a consequence of the acceleration mechanism taking place at the sources and subsequent escape from the sources themselves. 
Hence, they do not offer a direct constraint on the scaling of the diffusion coefficient with rigidity, which is one of the key predictions of the theory: only the purely secondary species can be exploited to this purpose. Regarding secondaries, an indication of an even more pronounced hardening in secondary species is also present, suggesting a transport origin for the feature~\citep{Genolini:2017dfb}.
Such spectral feature may be attributed, for instance, as discussed in \citet{Blasi:2012yr,Aloisio:2015rsa} (see also \citet{Farmer2004} for a pioneering prediction) to a transition between two different regimes: (i) the low-energy range where CR transport is expected to be dominated by self-confinement due to the generation of Alfvén waves via {\it streaming instability}; (ii) the high-energy range where CR confinement is expected to be dominated by scattering off pre-existing turbulent fluctuations (\textit{i.e.}, for which self-generation effects are not expected to play a relevant role).

\begin{figure}
  \centering
  \includegraphics[width=0.8\textwidth]{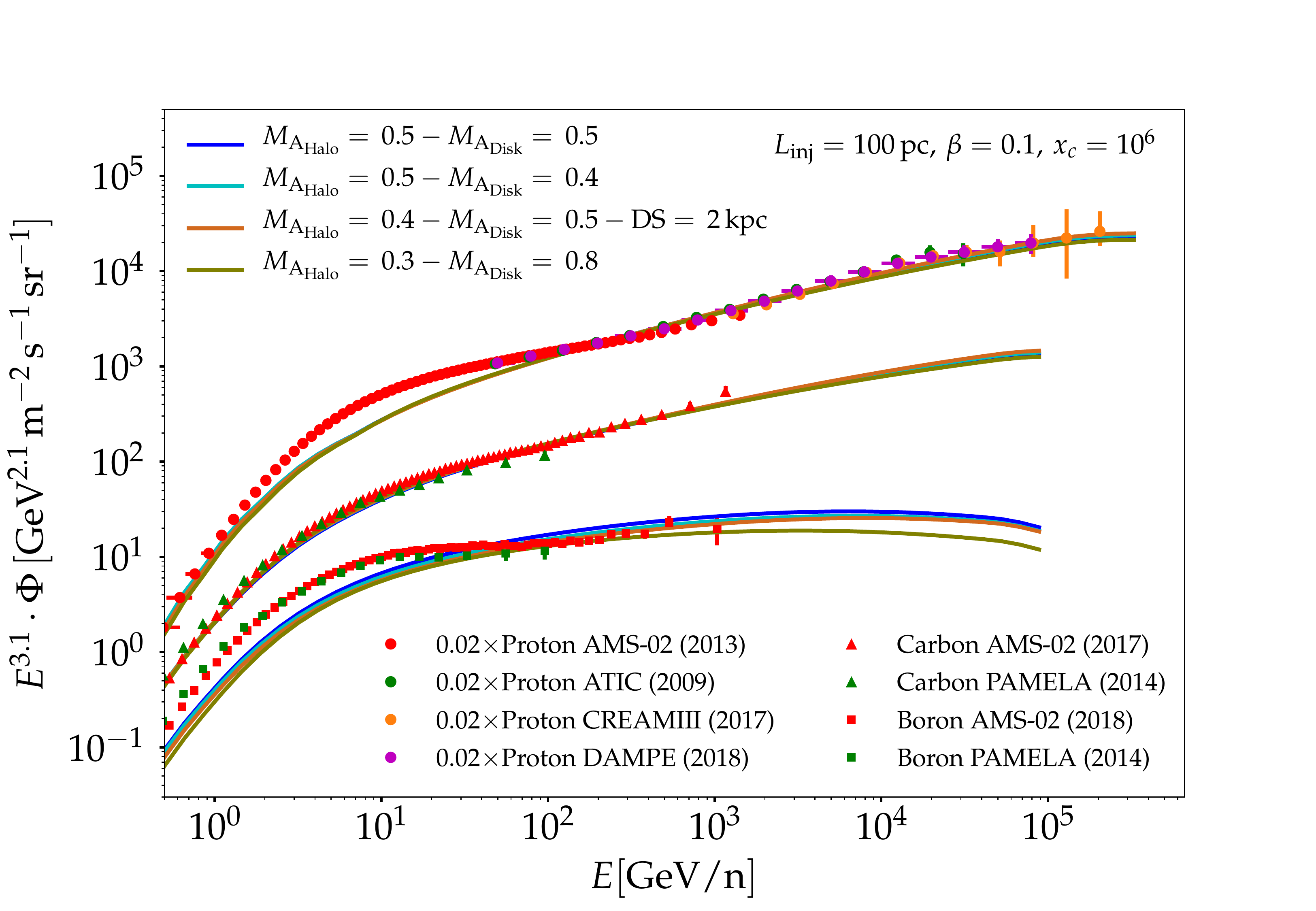}
\caption{\it {\bf Fluxes of H, B, C.} We plot the theoretical prediction for the Hydrogen, Carbon and Boron fluxes (obtained with an updated version of the {\tt DRAGON} code) for a few selected combinations of the parameters of interest. The primary injection spectrum is tuned to fit the data above 200 GeV. All high-energy data can be consistently reproduced within our theoretical framework. An extra confinement mechanism may be required to explain the low-energy excess.} 
  \label{fig:HBC}
\end{figure}

Motivated by these considerations, and given the aspects highlighted in the study of B/C, we aim at providing a comprehensive picture of the {\it high-energy portion of the spectrum}, above the aforementioned break, where the confinement due to scattering onto {\it isotropic fast-magnetosonic turbulence} should be the dominant physical mechanism (\textit{i.e.}, given on the one hand the lower impact of self-confinement and, on the other hand, the negligible role played by scattering on the pre-existing anisotropic Alfv\'enic cascade).
In the case of primary species, we aim at identifying a reasonable choice of the injection spectrum that correctly reproduce the data, given the degeneracy mentioned above.

In Figure \ref{fig:HBC} we show a particular realization that satisfies all the experimental constraints in the high-energy regime.
We show that we can consistently reproduce all the observed data above the $200 \, \mathrm{GV}$ spectral feature, by assuming a reasonable injection slope ($\gamma = 2.3$) and propagating the particles within the setup described above.  The ``excess'' at low energy cannot be reproduced within the framework discussed in the present work, and, given the considerations above, it strongly suggests the presence of another confinement mechanism, possibly associated to the self-generation of Alfv\'enic turbulence via streaming instability.

\section{Discussion and future prospects}\label{sec:discussion}

This paper is aimed at presenting the first comprehensive  study on the phenomenological implications of the theory describing cosmic-ray scattering onto magnetosonic fluctuations.
In this section we discuss potential caveats and future developments of the current work.

As a first discussion point, we want to argue on the potential impact of the anisotropic nature of cosmic-ray transport. In this paper, following the line of thought outlined for instance in \cite{Strong:2007nh} and adopted in most papers featuring a numerical description of cosmic-ray transport, we worked under the hypothesis of isotropic diffusion, assuming that the same scaling relations apply to parallel and perpendicular transport (see also \citealt{Evoli:2013lma}).
Within the current theoretical framework, this is formally correct only for values of $M_A \simeq 1$. In fact, in \citet{Yan:2007uc} the authors demonstrated that the perpendicular coefficients in this (weakly) non-linear extension of the QLT of scattering onto fast magnetosonic modes depend very strongly on the Alfvénic Mach number of the turbulence, exhibiting a $\propto M_{\mathrm{A}}^4$ scaling. 
However, many different mechanisms may lead to an effective isotropization of the diffusion tensor\footnote{For instance, the role of {\it compound diffusion}, resulting from the convolution of diffusion in the parallel and perpendicular directions with respect to the magnetic field line, has been studied in a series of papers, where, in particular, the role of {\it field line random walk} (FLRW) is found to be very important, especially for small turbulence perturbations~\citep{1966ApJ...146..480J, 1969ApJ...155..799J,
1969ApJ...155..777J, 2000ApJ...531.1067K, 2004A&A...420..821S, 2006ApJ...651..211W}.} and a commonly adopted assumption that has allowed to successfully reproduce all local observables is that CR transport is well described by an effective scalar coefficient. We also remark that the interaction of CRs with Alfv\'enic turbulence, which was shown to have a negligible impact on pitch-angle scattering, may play a role in this particular context. It can therefore in principle contribute to the perpendicular transport, and eventually to the isotropization of the diffusion tensor.
A careful assessment of this aspect is clearly well beyond the scope of the present work. In fact, it would require a full three-dimensional anisotropic treatment of CR diffusion and a careful modeling of the topology of the Galactic regular magnetic field.
However, in future studies, we will address in more detail the intrinsic anisotropic nature of CR transport within the theory presented here. We expect that the impact of a different scaling for the perpendicular transport may potentially be of some relevance as far as non-local observables --- $\gamma$-rays and radio waves for instance --- are concerned, especially in regions that feature values of $M_{\mathrm{A}}$ significantly smaller than $1$ (see \citet{Cerri:2017} for a pioneering study on the impact of anisotropic transport on non-local CR observables).

Another important aspect that potentially requires a dedicated study is the interplay with self-confinement due to Alfvénic turbulence originated by {\it CR-streaming instability}. As pointed out in \citet{Farmer2004,Blasi:2012yr}, this effect may dominate the low-energy confinement. As a consequence, the transition between a confinement regime dominated by scattering off self-generated turbulence and a regime dominated by scattering onto pre-exisiting MHD turbulence may be the origin of the spectral feature at $\simeq 200 \, \mathrm{GV}$ outlined in detail by the AMS-02 Collaboration in all the CR species.
On the other hand, we have shown that the relative importance of Alfvénic confinement progressively increases at high energy (Figure \ref{fig:coefficients_halo_AllMHD_vs_alfvenOnly} upper panel). This is due to the lower degree of anisotropy of the Alfvénic cascade at scales closer to the injection scale. Consequently, a spectral feature may be present in the high-energy spectrum, close to the PeV domain. A careful assessment of such a feature, its dependence on the environmental properties, and the potential of future experiments (such as LHAASO) to detect it, may constitute another very interesting future avenue in this research field. 

As a final discussion point, we mention the necessity to perform complementary observations and analyses aimed at highlighting the actual presence of magnetosonic fluctuations in the interstellar plasma. In this context, the statistical study of the Stokes parameters of the synchrotron-radiation polarization is a very promising technique. As recently demonstrated in \citet{2020NatAs...4.1001Z}, polarization analyses provide a unique opportunity to shed light on the plasma modes composition of the Galactic turbulence, and have led to a discovery of magnetosonic modes in the Cygnus X superbubble.

As a take-home message for this discussion, we want to emphasize the complementarity between different approaches. On the one hand, the arguments above outline the need of a dedicated effort from the experimental side, regarding direct measurements of local CR fluxes, aimed at detecting and characterizing spectral features over a wide energy range and with particular focus on the TeV - PeV domain. On the modeling side, we have emphasized the potential for a significant advance, aimed at analyzing the prediction of the theories in a realistic framework that takes into account the three-dimensional structure of the Galaxy, the topology of its magnetic field, and the properties of the interstellar medium. Both efforts are complemented by a research program directed towards analyzing the properties of interstellar turbulence. Thanks to the interplay among these developments, we may finally shed light on the long-standing puzzle of cosmic-ray confinement in the Galaxy.



\section{Summary}\label{sec:summary}

In this paper we have presented for the first time a comprehensive phenomenological study adopting a (weakly) non-linear extension of the quasi-linear theory of cosmic-ray scattering onto magneto-hydro-dynamic (MHD) fluctuations.

We have considered a state-of-the-art description of pitch-angle scattering associated to the various MHD cascades, \textit{i.e.}, decomposed into a (anisotropic) cascade of Alfv\'enic fluctuations, and slow and fast (isotropic) magnetosonic turbulence. We have studied the physical problem of the interaction of a charged, relativistic particle with such modes and, adopting the formalism developed in \citet{Yan:2004aq,Yan:2007uc}, we have computed the associated transport coefficients from first principles.

We identified a set of parameters that characterize the interstellar medium and have significant impact on our result (\textit{i.e.} the Alfv\'enic Mach number, the plasma $\beta$, and some parameters that describe the damping processes in different environments), and presented a complete phenomenological study of the dependence of the diffusion coefficients with respect to those parameters.

Then, we implemented the coefficients in the numerical framework of the {\tt DRAGON2} code, and tested the theory against current experimental data, with particular focus on the extremely accurate AMS-02 dataset. We found that the high-energy behaviour of the transport coefficients nicely matches the secondary-over-primary slope in that regime, and a reasonable range of the aforementioned parameters allowed us to reproduce the correct normalization as well, without invoking any {\it ad hoc} tuning. Overall, we found a natural and reasonable agreement with all CR channels within a reasonable choice of both the ISM parameters governing the transport process, and other parameters (\textit{e.g.} injection slope) that characterize our setup. 

The theory is therefore adequate to describe the microphysics of Galactic CR confinement in the high-energy domain, in particular above the $200 \, \mathrm{GeV}$ feature highlighted in all primary and secondary species by the AMS-02 Collaboration. On the other hand, we confirm that the pitch-angle scattering on pre-existing Alfv\'enic turbulence can not provide a satisfactory description of CR confinement: in fact, the highly anisotropic Alfv\'enic cascade turns out to be extremely inefficient in scattering CRs of energies $\lesssim 100\, \mathrm{TeV}$. Our work strongly suggests that the interpretation of AMS-02 data in terms of pitch-angle scattering onto turbulent fluctuations naively described in terms of a Kolmogorov-like isotropic spectrum cannot be considered satisfactory, and a more accurate description of interstellar turbulence has to be considered.

The behaviour of CR observables below $200 \, \mathrm{GV}$ cannot be reproduced within our framework. The steeper spectrum observed by AMS-02 below that energy seems to require additional physical effects. The self-confinement due to self-generated Alfv\'enic fluctuations via {\it CR-streaming instability} seems a good candidate in this context. We postpone to a forthcoming study a full combined treatment of this process within our formalism and our setup.

\section*{Acknowledgments}

We are grateful to Carmelo Evoli and Huirong Yan for many inspiring discussions.
OF, SSC, and PDT warmly thank the generous hospitality of the Instituto de F\'isica Te\'orica (IFT) in Madrid.
The work of DG has received financial support through the Postdoctoral Junior Leader Fellowship Programme from la Caixa Banking Foundation (grant n. LCF/BQ/LI18/11630014).
DG was also supported by the Spanish Agencia Estatal de Investigaci\'{o}n through the grants PGC2018-095161-B-I00, IFT Centro de Excelencia Severo Ochoa SEV-2016-0597, and Red Consolider MultiDark FPA2017-90566-REDC.
OF was supported by the University of Siena with a joint doctoral degree with the Autonomous University of Madrid.
SSC was supported by the Max-Planck/Princeton Center for Plasma Physics (NSF grant PHY-1804048).
PDT was supported by the University of Bari and INFN, Sezione di Bari.
SG acknowledges support from Agence Nationale de la Recherche (grant ANR- 17-CE31-0014).

\section*{Data availability}

No new data were generated or analysed in support of this research.

A new scientific code to compute the diffusion coefficients is available in a repository and can be accessed via the DOI link \url{https://doi.org/10.5281/zenodo.4250807}.

\clearpage
\appendix
\section{Pitch-angle diffusion coefficient for MHD turbulence}\label{app:D_mumu_for_MHD}

In this appendix we briefly review the calculations carried out in \citet{Yan:2007uc} to compute the relative contributions from each of the MHD modes to the spatial diffusion coefficient. In particular, in \citet{Yan:2007uc} the authors mostly implement the case of trans-alfv\'enic turbulence $(M_{\rm A} \simeq 1)$, whereas {here we consider a broader range of Alfv\'enic Mach number, pertaining also to sub- and super-Alfv\'enic regimes (\textit{i.e.}, roughly within the range $0.1\lesssim M_{\rm A} \lesssim 2$)}. 

The starting point is Equation \eqref{eq:pitch-angle_diffusion_general} {for the pitch-angle scattering rate of a charged particle in turbulent fluctuations, that we report here for convenience:} 
\begin{equation}\label{eq:pitch-angle_diffusion_general_APPENDIX}
{    
D_{\mu \mu} = \Omega^2 (1 - \mu^2) \int d^3 \bm{k} \sum^{+\infty}_{n=-\infty} R_n (k_{\parallel} v_{\parallel} - \omega + n \Omega) \left[ \frac{n^2 J_n^2(z)}{z^2}I^{\rm A} (\bm{k}) + \frac{k^2_{\parallel}}{k^2} J'^{2}_{n}(z) I^{\rm M} (\bm{k}) \right],
}
\end{equation}
where we remind the reader that $\Omega=qB_0/m \gamma c$ is the particle's gyro-frequency, $\mu=v_\|/|\bm{v}|=\cos\theta$ its pitch angle ($\theta$ being the angle between the particle's velocity $\bm{v}$ and the background magnetic field $\bm{B}_0$), $\bm{k}$ and $\omega$ are the fluctuations' wave-vector and frequency, respectively, and $I(\bm{k})$ their turbulent power spectrum at scales $\sim k^{-1}$ (which is modified by a combination of the Bessel's functions $J_n(k_\perp r_{\rm L})$, as effectively seen through a particle's gyro-motion whose Larmor radius is $r_{\rm L}$, and that scatters via a resonance-like function $R_n$).

{To model the turbulent fluctuations of the magnetic field and of the fluid velocity at MHD scales, $\delta B$ and $\delta u$, respectively,} we follow the prescription given in \citet{PhysRevLett.89.281102} {for their correlation functions}:
\begin{subequations}\label{eq:turbulent_spectra_general}
\begin{align}
    \langle \delta B_i (\bm{k}) \cdot \delta B^*_j (\bm{k'}) \rangle / B^2_0\, & \,=\, \delta^3(\bm{k} - \bm{k'}) \, {\cal M}_{ij}(\bm{k}) \label{eq:correlation_general_BB}\\
    \langle \delta u_i (\bm{k}) \cdot \delta B^*_j (\bm{k'}) \rangle / v_A B_0\, & \,=\, \delta^3(\bm{k} - \bm{k'}) \, {\cal C}_{ij}(\bm{k}) \label{eq:correlation_general_vB}\\
    \langle \delta u_i (\bm{k}) \cdot \delta u^*_j (\bm{k'}) \rangle / v_A^2\, & \,=\, \delta^3(\bm{k} - \bm{k'}) \, {\cal K}_{ij}(\bm{k}),\label{eq:correlation_general_vv}
\end{align}
\end{subequations}
where the {indices} $i,j = 1,2,3$ {represent} the different components of the {fluctuation vector}, and the $\langle \, \rangle$ operator indicates the average over a phase-space ensemble~\citep{doi:10.1143/JPSJ.12.570}.
{These correlation functions are related to the energy density of the fluctuations, \textit{e.g.}, $\langle \delta B(\bm{x}) \delta B^*(\bm{x}) \rangle$ for magnetic-field fluctuations. In fact,} in terms of their Fourier components, $\delta B(\bm{x}) = \int d^3 \bm{k} \, e^{-i \bm{k \cdot x} } \, \delta B(\bm{k})$ and $\delta B^*(\bm{x}) = \int d^3 \bm{k} \, e^{i \bm{k \cdot x} } \, \delta B^*(\bm{k})${, the fluctuations' energy density can be written as}
\begin{align}\label{eq:fluctuation_energy_density}
    \langle \delta B(\bm{x})^2 \rangle &= \sum_{i,j} \int d^3 \bm{k} \int d^3 \bm{k'} \, e^{-i \bm{(k - k') \cdot x}} \, \langle \delta B_i(\bm{k}) \cdot \delta B_j^*(\bm{k'}) \rangle \\
    &= B_0^2 \cdot \sum_{i,j} \int d^3 \bm{k} \; {\cal M}_{ij}(\bm{k}),
\end{align}
such that the integral of the normalized fluctuation spectrum {over wave-numbers} gives the spatial energy density. This is in {agreement} with \citet{1975RvGSP..13..547V} (their Equation (32)).
{The spectrum of a given turbulent field is then obtained as the trace of the correlation tensor of its fluctuations. For instance, the trace of ${\cal M}_{ij}$ provides the magnetic-field turbulent spectrum:} $\sum_{i=j} {\cal M}_{ij} = I^{\rm A,S,F}$, where ${\rm A}$ labels the Alfv\'en mode, and ${\rm S}$, ${\rm F}$ the slow and fast magnetosonic modes, respectively. In what follows, only the magnetic-field fluctuations {and their correlation tensor in \eqref{eq:correlation_general_BB} will enter the calculations.}

For what concerns the explicit form of the magnetic-field fluctuations' correlation tensor ${\cal M}_{ij}$, we will make use of the expressions outlined in \citet{Cho:2001hf}, which were obtained via numerical simulations in the trans-Alfv\'enic regime $M_{\rm A}\simeq1$.
However, as mentioned above, {in this work} we consider {turbulent regimes that span} from the sub-Alfv\'enic $(M_{\rm A} < 1)$ to the super-Alfv\'enic $(M_{\rm A} > 1)$ case.
Therefore, the general correlation tensor (and the corresponding turbulent spectrum) of the Alfv\'enic and slow cascades that will be considered here must include an extra scaling with the Alfv\'enic Mach number $M_{\rm A}$ (a scaling that also depends whether we are in the sub-Alfv\'enic or in the super-Alfv\'enic case, as outlined in \citet{Lazarian:2020cms}, and from which the usual GS95 scaling~\citep{GS1995ApJ...438..763G} is anyway recovered in the trans-Alfv\'enic limit, $M_{\rm A}\sim1$). 
By taking into account these generalizations, the correlation tensors pertaining to the three MHD modes scale as follows:

\begin{align}
\label{eq:alfven_scaling_sub}
&\mathcal{M}_{ij}^{\rm A(S),sub} = C^{\rm A(S),sub}_a I_{ij} k_{\perp}^{-10/3} \cdot \exp{\left( -\frac{L^{1/3} |k_{\parallel}|}{M_{\rm A}^{4/3} \, k_{\perp}^{2/3}} \right)} \qquad \qquad \; \; (M_{\mathrm{A}} \leq 1) \\
\label{eq:alfven_scaling_super}
&\mathcal{M}_{ij}^{\rm A(S),super} = C^{\rm A(S),super}_a I_{ij} k_{\perp}^{-10/3} \cdot \exp{\left( -\frac{L^{1/3} |k_{\parallel}|}{M_{\rm A} \, k_{\perp}^{2/3}} \right)} \qquad \qquad (M_{\mathrm{A}} > 1) \\
\label{eq:fast_scalings}
&\mathcal{M}_{ij}^{\rm F} = C^{\rm F}_a J_{ij} k^{-7/2},
\end{align}
where $C_a$ are normalization constants, and parallel ($\parallel$) and perpendicular ($\perp$) here are defined with respect to the background magnetic field, $\bm{B}_0$. 
The tensors $I_{ij} = \delta_{ij} - k_i k_j / k_{\perp}^2$ and $J_{ij} = k_i k_j / k_{\perp}^2$ are 2D tensors defined in the sub-space perpendicular to the background magnetic field\footnote{{If $\mathcal{I}_{ij}=\delta_{ij}-k_ik_j/k_\perp^2$ and $\mathcal{J}_{ij}=k_ik_j/k_\perp^2$ are the 3D version of $I_{ij}$ and $J_{ij}$ defined for any $i,j=1,2,3$ index, then the 2D version can be generally written as $I_{ij}=\mathcal{T}_{ik}\mathcal{I}_{kl}\mathcal{T}_{lj}$ and $J_{ij}=\mathcal{T}_{ik}\mathcal{J}_{kl}\mathcal{T}_{lj}$, with $\mathcal{T}_{ij}=\delta_{ij}-B_{0,i}B_{0,j}/B_0^2$ being the projecting operator onto the plane perpendicular to $\bm{B}_0$.}} (\textit{e.g.}, if $\bm{B}_0$ is along $z$, then $I_{ij}$ and $J_{ij}$ above are defined in the $xy$-plane, and are zero if $i,j=z$).
Within the plane perpendicular to $\bm{B}_0$, $I_{ij}$ and $J_{ij}$ are indeed projecting operators working as expected for the polarization of the Alfv\'en, slow and fast modes: $I_{ij}$ projects onto the direction perpendicular to $\bm{k}_{\perp}$, whereas $J_{ij}$ projects onto the direction parallel to it.
As {an additional remark}, we {point out} that {the above} scalings are the 3D extensions of the 1D spectra found in \citet{ChoPhysRevLett.88.245001}.

Finally, in order to determine the normalization constants $C_a$, we require that the energy of the turbulent fluctuations obtained by their correlation tensor (\textit{i.e.}, $\langle \delta B(\bm{x})^2 \rangle$ from Equation \eqref{eq:fluctuation_energy_density}) matches the root-mean-square value of the fluctuations at the injection scale $L$ (\textit{i.e.}, $\delta B_{\mathrm{rms}}^2 \equiv \langle \delta B^2 \rangle_L$):
\begin{equation}\label{eq:normalization_MHD_modes}
\langle \delta B(\bm{x})^2 \rangle\,
\equiv\, 
B_0^2\, \sum_{i,j} \int d^3 \bm{k}\, {\cal M}_{ij}(\bm{k})\,
\overset{!}{=}\,
\delta B_{\mathrm{rms}}^2\, 
\equiv\, 
\langle \delta B^2 \rangle_L\,,
\end{equation}
where $\langle\delta B^2\rangle_L$ is related to the (outer-scale) Alfv\'enic Mach number $M_{\rm A}$ by $\langle\delta B^2\rangle_L/B_0^2\approx M_{\rm A}^2$.

\subsection{$D_{\mu\mu}$ from Alfv\'en modes}

In this Section, we specialize to the case of a cascade of Alfv\'enic fluctuations, explicitly providing the steps of the calculation leading to the associated pitch-angle scattering rate, $D_{\mu\mu}^{\rm A}$, in the relevant regimes.

\subsubsection{Normalization coefficient}

To get the normalization {coefficient $C_a^{\rm A}$} for the {Alfvénic cases}, we make use of Equation \eqref{eq:normalization_MHD_modes} with the spectrum \eqref{eq:alfven_scaling_sub} {or \eqref{eq:alfven_scaling_super} for the sub- or super-Alfv\'enic regime, respectively}, where $\sum_{i=j} I_{ij} = 1$. {Moreover, since Alfv\'enic fluctuations} are anisotropic, it is convenient to write the integral decomposing it as $\int d^3 \bm{k} = \int^{+ \infty}_{L^{-1}} k_{\perp} d k_{\perp} \int^{+ \infty}_{- \infty} d k_{\parallel} \int^{2 \pi}_{0} d \phi$.

\vspace{0.3cm}
\hspace{-0.6cm}\textit{\textbf{Sub- and trans-Alfv\'enic regime ($M_{\rm A} \leq 1$)}}. When dealing with sub-Alfv\'enic turbulence, the cascade of fluctuations at scales immediately below the injection scale belongs to the weak-turbulence regime. This means that, initially, fluctuations develop a $E(k_{\perp}) \sim k_{\perp}^{-2}$ spectrum in the direction perpendicular to $B_0$, while there is no turbulent cascade along the magnetic-field lines, $E(k_\|)=E(k_L)=\mathrm{cst}$. However, this weak-turbulence regime can be sustained only for a limited range of (perpendicular) scales, $[L^{-1}, \ell_{\mathrm{tr}}^{-1}]$, as the critical-balance condition will be anyway achieved at a scale $\ell_{\mathrm{tr}} \sim L M_{\mathrm{A}}^2$ that determines the transition to the strong-turbulence regime~\citep{GS1995ApJ...438..763G}.
At perpendicular scales $\lambda_\perp\leq\ell_{\mathrm{tr}}$, the cascade follows the modified GS95 spectrum in \eqref{eq:alfven_scaling_sub}. Therefore, to obtain the normalization constant $C_a^{\rm A,sub}$, we now use the fact that the integral of the magnetic-field fluctuations' correlation tensor should match the energy of the fluctuations at the transition scale $\ell_{\mathrm{tr}}$, \textit{i.e.}, $\int d^3 \bm{k}\, {\cal M}_{ij}^{\rm A,sub}(\bm{k}) =\langle\delta B^2\rangle_{\ell_{\rm tr}}/B_0^2$. Also, since the parallel scale does not evolve in the weak-turbulence regime, we remind the reader that the parallel wavelength $\lambda_{\|,{\rm tr}}$ corresponding to a turbulent eddy of perpendicular size $\lambda_\perp=\ell_{\rm tr}$ is still the injection scale, \textit{i.e.} $\lambda_{\|,{\rm tr}}=L$. As a result, the exponential function that describes surfaces of constant energy in $(k_\perp,k_\|)$ space still contains the outer-scale factor, $L^{1/3}$, as for the trans-Alfv\'enic limit, $M_{\rm A}=1$.
The equation {that determines $C_a^{\rm A,sub}$} is therefore:

\begin{equation*}
\begin{aligned}
    &C^{\rm A,sub}_a \cdot 2 \pi \int^{k_{\perp, \mathrm{max}}}_{\ell_{\mathrm{tr}}^{-1}} k_{\perp} d k_{\perp} \int_{_{  \left[ -k_{\parallel, \mathrm{max}}, -L^{-1} \right] \cup \left[ L^{-1}, k_{\parallel, \mathrm{max}} \right] }} d k_{\parallel} \, k_{\perp}^{-10/3} \cdot \exp{\left( -\frac{L^{1/3} |k_{\parallel}|}{M_{\rm A}^{4/3} \, k_{\perp}^{2/3}} \right)} \approx \\
    \approx \;
    & C^{\rm A,sub}_a \cdot 2 \pi \int^{+ \infty}_{\ell_{\mathrm{tr}}^{-1}} k_{\perp} d k_{\perp} \int^{+ \infty}_{- \infty} d k_{\parallel} \, k_{\perp}^{-10/3} \cdot \exp{\left( -\frac{L^{1/3} |k_{\parallel}|}{M_{\rm A}^{4/3} \, k_{\perp}^{2/3}} \right)}
    \,\overset{!}{=}\, 
    \frac{\langle \delta B^2 \rangle_{\ell_{\mathrm{tr}}}}{B_0^2}.
\end{aligned}
\end{equation*}

The above approximations in the limits of integration involve both the cutoff and the injection wave-number scales: 
(i) the former corresponds to the cascade cutoff scales $(k_{\perp, \mathrm{max}}, k_{\parallel, \mathrm{max}})$, and letting them approach infinity does not lead to any appreciable modification. Indeed, the perpendicular spectrum is soft enough $(E(k_{\perp}) \sim k_{\perp}^{-10/3})$ that the large wave-numbers carry very little turbulent power. 
In particular, this is true for the parallel spectrum, since the GS95 critical-balance relation implies an even softer spectrum versus $k_\parallel$. (ii) As far as the low-$k_\parallel$ limit of integration is concerned, considering the proper injection scale ($k_{\parallel, \mathrm{min}} \sim L^{-1}$) introduces a correction factor $1/e$ in the normalization constant. This correction only affects Alfvèn and slow modes, that will be found to be anyway strongly subdominant in shaping the cosmic-ray diffusion coefficient, therefore, for the sake of simplicity, we neglect it. Notice, however, that we will use this approximation only for the normalization constant, while the correct wave-number range is considered when calculating $D_{\mu \mu}$, thus not affecting the resulting slopes of the diffusion coefficient.

Solving the integrals, the left-hand side of the above equation yields $C_a^{\rm A,sub} \, 4 \pi \cdot \frac{3 M_{\rm A}^{4/3} \, \ell_{\mathrm{tr}}^{2/3}}{2 L^{1/3}}$.
{Then,} taking into account the scaling $\ell_{\mathrm{tr}} \sim L M_{\rm A}^2$ {for the transition scale}, we can obtain the normalization in terms of the injection scale $L$:
\[
    C_a^{\rm A,sub}\, 4 \pi \cdot \frac{3}{2}\, L^{1/3} \, M_{\rm A}^{8/3} 
    \,\overset{!}{=}\, 
    \frac{\langle \delta B^2 \rangle_{\ell_{\mathrm{tr}}}}{\langle \delta B^2 \rangle_{L}} \cdot \frac{\langle \delta B^2 \rangle_{L}}{B_0^2} \approx M_{\rm A}^2 \cdot M_{\rm A}^2 = M_{\rm A}^4,
\]
where we have used the scaling of weak turbulence for the fluctuations, $\delta B_\lambda\sim\lambda_\perp^{1/2}$, to substitute $\langle \delta B^2 \rangle_{\ell_{\mathrm{tr}}}/\langle \delta B^2 \rangle_{L} = \ell_{\rm tr}/L \approx M_{\rm A}^2$, and $\langle \delta B^2 \rangle_{L}/B_0\equiv M_{\rm A}^2$.  

In conclusion, $C_a^{\rm A,sub} = M_{\rm A}^{4/3} \, L^{-1/3}/6 \pi$ and the {correlation tensor of the magnetic-field fluctuations for} the Alfv\'en mode in the sub-Alfv\'enic {(or trans-Alfv\'enic) regime} is:
\begin{equation}\label{eq:alfven_spectrum_normalized}
    {\cal M}^{\rm A,sub}_{ij} = \frac{M_{\rm A}^{4/3} \, L^{-1/3}}{6 \pi}\, I_{ij}\, k_{\perp}^{-10/3} \cdot \exp\left(- \frac{L^{1/3} k_{\parallel}}{M_{\rm A}^{4/3} \, k_{\perp}^{2/3}} \right).
\end{equation}

\vspace{0.3cm}
\hspace{-0.6cm}\textit{\textbf{Super-alfv\'enic case: $\bm{M_{\mathrm{A}} > 1}$}}. Conversely to what happens in the sub-Alfv\'enic case, when the injected fluctuations are super-Alfv\'enic, the corresponding turbulent cascade at scales immediately below the injection scale $L$ is ``hydro-dynamical'' in nature, \textit{i.e.}, isotropic with a spectrum $E(k)\sim k^{-5/3}$. 
This hydrodynamic-like behaviour is, again, sustained only within a limited range of scales, $[L^{-1},\ell_{\rm A}^{-1}]$, as the critical-balance condition will be eventually met at the Alfv\'en scale $\ell_{\mathrm{A}} \sim L M_{\rm A}^{-3}$~\citep{Lazarian:2020cms}. At scales $\lambda\leq\ell_{\rm A}$ the turbulent cascade thus becomes anisotropic with respect to the magnetic-field direction, and follows the modified GS95 spectrum in \eqref{eq:alfven_scaling_super}.
Following the same reasoning of the sub-Alfv\'enic case, the equation for $C_a^{\rm A,super}$ reads as 
\begin{equation*}
    C^{\rm A,super}_a \cdot 2 \pi \int^{+ \infty}_{\ell_{\mathrm{A}}^{-1}} k_{\perp} d k_{\perp} \int^{+ \infty}_{- \infty} d k_{\parallel} \, k_{\perp}^{-10/3} \cdot \exp{\left( -\frac{L^{1/3} |k_{\parallel}|}{M_{\rm A} \, k_{\perp}^{2/3}} \right)} 
    \,\overset{!}{=}\, 
    \frac{\langle \delta B^2 \rangle_{\ell_{\mathrm{A}}}}{B_0^2}.
\end{equation*}
{By explicitly} solving the integral and taking into account the scaling $\ell_{\mathrm{A}} \sim L M_{\mathrm{A}}^{-3}$, {one obtains}:
\[
    4 \pi C_a^{\rm A,super} \cdot \frac{3}{2} L^{1/3} \, M_{\rm A}^{-1} 
    \,\overset{!}{=}\,
    \frac{\langle \delta B^2 \rangle_{\ell_{\mathrm{A}}}}{\langle \delta B^2 \rangle_{L}} \cdot \frac{\langle \delta B^2 \rangle_{L}}{B_0^2} \approx M_{\rm A}^{-2} \cdot M_{\rm A}^2,
\]
where we have used the Kolmogorov-like scaling for the turbulent fluctuations, $\delta B_\lambda \sim \lambda^{1/3}$, to substitute $\langle \delta B^2 \rangle_{\ell_{\mathrm{A}}}/\langle \delta B^2 \rangle_L = (\ell_{\rm A}/L)^{2/3} \approx M_{\rm A}^{-2}$, and, again, $\langle \delta B^2 \rangle_{L}/B_0^2 = M_{\rm A}^2$ by definition.

In conclusion, $C_a^A = M_{\rm A} \, L^{-1/3}/6 \pi$, and the {correlation tensor of the magnetic-field fluctuations for the Alfv\'en mode in the super-Alfv\'enic regime is:}
\begin{equation}\label{eq:alfven_spectrum_normalized_super}
    {\cal M}^{\rm A,super}_{ij} = \frac{M_{\rm A} \, L^{-1/3}}{6 \pi}\, I_{ij}\, k_{\perp}^{-10/3} \cdot \exp\left(- \frac{L^{1/3} k_{\parallel}}{M_{\rm A} \, k_{\perp}^{2/3}} \right).
\end{equation}

\subsubsection{Resonance function}
In this work, we are adopting the resonance function, $R_n$, described in \citet{Yan:2007uc}. Such function includes the broadening of the resonant scattering wave-number due {finite-amplitude corrections in the magnetic-field strength\footnote{{An effect that is consistent with the inclusion in this scattering theory of the Landau-type wave-particle interaction usually referred to as \textit{transit-time damping} (TTD).}}}:
\begin{equation*}\label{eq:resonant_function}
    R_n (k_{\parallel} v_{\parallel} - \omega + n \Omega) = \frac{\sqrt{\pi}}{|k_{\parallel}| v_{\perp} M_{\rm A}^{1/2}} \cdot \exp \left(- \frac{(k_{\parallel} v \mu - \omega + n\Omega)^2}{k^2_{\parallel} v^2 (1 - \mu^2) M_{\rm A}} \right),
\end{equation*}
where {we recall the reader that the above expression reduces to the usual Dirac $\delta$-function in the limit of vanishing fluctuations amplitude, $M_{\rm A}\to0$}.

{Within the present approximations, Alfv\'en modes can scatter CRs only via $n \neq 0$ gyro-resonance interactions, while the $n=0$ Landau-damping interaction is neglected.} 
Also, {we consider low-frequency, non-relativistic MHD turbulence, \textit{i.e.}, turbulent fluctuations within a range of frequencies $\omega$ and wave-numbers $\bm{k}$ such that their frequency is much smaller than the particles' gyro-frequency, $\omega\ll\Omega$, and their phase velocity is much smaller than the speed of light, $v_{\rm ph}\sim\omega/k\ll c$.
In this limit, since cosmic particles' are relativistic (\textit{i.e.}, their velocity is typically $v\approx c$), one can neglect the fluctuation frequency $\omega$ in the argument of the resonance function: $k_{\parallel} v \mu - \omega + n\Omega \simeq k_{\parallel} v \mu + n\Omega$.

Taking these considerations into account and rearranging the argument of the exponential, the resonance function that will be adopted for scattering on Alfv\'enic fluctuations reads
\begin{align}
    R_n (k_{\parallel} v_{\parallel} - \omega + n \Omega) =\, &\, \frac{\sqrt{\pi}}{|k_{\parallel}| v_{\perp} M_{\mathrm{A}}^{1/2}} \cdot \exp \left(- \frac{\left( \mu + \frac{n}{x_{\parallel} R} \right)^2}{(1 - \mu^2) M_{\mathrm{A}}} \right) \equiv \frac{\sqrt{\pi}}{|k_{\parallel}| v_{\perp} M_{\mathrm{A}}^{1/2}} \cdot E_n\nonumber\\
    =\, & \,\frac{\sqrt{\pi\,}\,\Omega^{-1}}{|x_\|| R\, M_{\mathrm{A}}^{1/2}} \cdot E_n\,,
\end{align}
where we have defined $R \equiv v/(\Omega L) = (1-\mu^2)^{-1/2}r_{\rm L}/L$, with $r_{\rm L}=v_\perp/\Omega$ the cosmic particle's Larmor radius, and $x_{\|,\perp} \equiv k_{\|,\perp}L$.
}

\subsubsection{Pitch-angle coefficient}\label{subsubsec:pitch-angle_alfven}
To finally calculate the contribution from the Alfv\'en modes to the pitch-angle diffusion coefficient, we now make use of the spectra in Equations \eqref{eq:alfven_spectrum_normalized} and \eqref{eq:alfven_spectrum_normalized_super} in the following expression:
\begin{equation}\label{eq:pitch-angle_diffusion_alfven_generic}
    D_{\mu \mu}^{\rm A} = \Omega^2 (1 - \mu^2) \int d^3 \bm{k} \sum^{+\infty}_{n=-\infty} \frac{\sqrt{\pi}}{|k_{\parallel}| v_{\perp} M_{\rm A}^{1/2}} \cdot E_n \left[ I^A (\bm{k}) \frac{n^2 J_n^2(z)}{z^2} \right].
\end{equation}

\vspace{0.3cm}
\hspace{-0.6cm}\textit{\textbf{$\bm{M_{\mathrm{A}} \leq 1}$}}. Using the dimensionless quantities described above, the expression for the pitch-angle scattering rate on Alfv\'enic fluctuations in the $M_{\rm A}\leq1$ regime reads:
\begin{align}\label{eq:pitch-angle_diffusion_alfven}
    D_{\mu \mu}^{\rm A} =\, &\, \frac{\sqrt{\pi} v \sqrt{1 - \mu^2\,} \, M_{\rm A}^{5/6}}{3 R^2 L} \int d x_{\perp} \int d x_{\parallel} \sum^{+ \infty}_{n=- \infty} \frac{n^2 J_n^2(z)}{z^2} \cdot \frac{x_{\perp}^{-7/3}}{|x_{\parallel}|} \cdot \exp \left(- \frac{x_{\parallel}}{M_{\rm A}^{4/3} \, x_{\perp}^{2/3}} \right) E_n\nonumber\\
    =\, & \,\frac{\sqrt{\pi}\,\Omega\, M_{\rm A}^{5/6}}{3 R} \int d x_{\perp} \int d x_{\parallel} \sum^{+ \infty}_{n=- \infty} \frac{n^2 J_n^2(z)}{z^2} \cdot \frac{x_{\perp}^{-7/3}}{|x_{\parallel}|} \cdot \exp \left(- \frac{x_{\parallel}}{M_{\rm A}^{4/3} \, x_{\perp}^{2/3}} \right) E_n, 
\end{align}
where now $z \equiv x_{\perp} R (1 - \mu^2)^{1/2}$. For Alfv\'en modes, $n \neq 0$, and we verified that the $n=\pm 1$ functions give the dominant contribution, so that $D_{\mu\mu}^{\rm A}\approx D^{{\rm A}, n=1}_{\mu \mu} + D^{{\rm A}, n=-1}_{\mu \mu}$. Using the property $J_{-n} (z) = (-1)^n J_n(z)$, from which it follows $J^2_{-n} (z) = J^2_n (z)$, we finally get:
\begin{equation}\label{eq:pitch-angle_diffusion_alfven_arranged}
    D^{{\rm A, sub}}_{\mu \mu} = \frac{2 \sqrt{\pi}\,\Omega\, M_{\rm A}^{5/6}}{3 R} \int d x_{\perp} \int d x_{\parallel} \frac{J_1^2(z)}{z^2} \cdot \frac{x_{\perp}^{-7/3}}{x_{\parallel}} \cdot \exp \left( - \frac{x_{\parallel}}{M_{\rm A}^{4/3} \, x_{\perp}^{2/3}} \right)(E^{+} + E^{-}) \\
\end{equation}
where $E^{+} \equiv E^{+}_{n=1} = \exp \left( - \frac{\left( \mu + \frac{1}{x_{\parallel} R} \right)^2}{(1 - \mu^2) M_{\rm A}} \right)$, $E^{-} \equiv E^{-}_{n=-1} = \exp \left( - \frac{\left( \mu - \frac{1}{x_{\parallel} R} \right)^2}{(1 - \mu^2) M_{\rm A}} \right)$ and the factor $2$ comes from taking the integral only on $x_{\parallel}>0$.

The lower boundary of integration can be found reminding that we integrate the GS95 spectrum from the scale where the critical balance is reached. For this $M_{\mathrm{A}} \leq 1$ case, we have seen that, up to the transition scale, the cascade evolves only in the direction perpendicular to the magnetic field. Therefore, we can write:
\begin{equation*}
\begin{aligned}
    &k_{\perp, \mathrm{min}} \ell_{\mathrm{tr}} = k_{\perp, \mathrm{min}} \cdot \left( \frac{\ell_{\mathrm{tr}}}{L} \right) L \overset{!}{=} 1 \quad \Rightarrow \quad x_{\perp, \mathrm{min}} = \frac{1}{\left( \frac{\ell_{\mathrm{tr}}}{L} \right)} \approx M_{\mathrm{A}}^{-2} \\
    &x_{\parallel, \mathrm{min}} = 1,
\end{aligned}
\end{equation*}
where we denoted with $\ell_{\mathrm{tr}}$ the scale where the turbulence becomes of GS95 type.

\vspace{0.3cm}
\hspace{-0.6cm}\textit{\textbf{$\bm{M_{\mathrm{A}} > 1}$}}. Following the same steps as for the $M_{\rm A} \leq 1$ case, we eventually obtain the following expression:
\begin{equation}\label{eq:pitch-angle_diffusion_alfven_arranged_super}
    D^{{\rm A,super}}_{\mu \mu} = \frac{2 \sqrt{\pi}\,\Omega\, M_{\rm A}^{1/2}}{3 R} \int d x_{\perp} \int d x_{\parallel} \frac{J_1^2(z)}{z^2} \cdot \frac{x_{\perp}^{-7/3}}{x_{\parallel}} \cdot \exp \left( - \frac{x_{\parallel}}{M_{\rm A} \, x_{\perp}^{2/3}} \right)(E^{+} + E^{-})\,.
\end{equation}

In this case, the lower boundary for the integration can be obtained considering that the cascade evolves isotropically until the transition scale $\ell_{\mathrm{A}}$ is reached. Hence, we obtain:
\begin{equation*}
\begin{aligned}
    &k_{\perp, \mathrm{min}} \ell_{\mathrm{A}} = k_{\perp, \mathrm{min}} \cdot \left( \frac{\ell_{\mathrm{A}}}{L} \right) L \overset{!}{=} 1 \quad \Rightarrow \quad x_{\perp, \mathrm{min}} = \frac{1}{\left( \frac{\ell_{\mathrm{A}}}{L} \right)} \approx M_{\mathrm{A}}^{3} \\
    &x_{\parallel, \mathrm{min}} \approx M_{\mathrm{A}}^{3}.
\end{aligned}
\end{equation*}

To evaluate the upper boundary of the integrals, we do not treat the two regimes separately and assume that Alfv\'en modes do not undergo significant damping and therefore the cascade proceeds up to the dissipation scale. Equivalently, we will truncate the integrals at a wave-number much larger than the inverse of the Larmor radius of the less energetic particle, $k_{\perp} \gg r_L^{-1}\big|_{E_{\mathrm{min}}}$. In practice, we will consider two order of magnitudes larger than that quantity. Since we are considering particles with energy as low as $10^{-2} \, \mathrm{GeV}$, with a Larmor radius of $r_L \simeq 3.37 \cdot 10^{12} \, \mathrm{cm} \left( \frac{p = 10^{-2} \mathrm{GeV}}{\mathrm{GeV}} \right) \left( \frac{10^{-6} \, \mathrm{G}}{B} \right) = 3.37 \cdot 10^{10} \, \mathrm{cm}$, this corresponds to $k_{\perp, \mathrm{max}} = 10^2 \cdot (3.37 \cdot 10^{10} \, \mathrm{cm})^{-1} = 3 \cdot 10^{-9} \, \mathrm{cm}^{-1}$. Also, according to the findings of the GS95 theory, $k_{\parallel} \propto k_{\perp}^{2/3}$. 

In conclusion, the upper bounds for the integrals are:
\begin{equation}\label{eq:upper_bound_alfven_integral}
    x_{\perp, \mathrm{max}} = 3 \cdot 10^{-9} \cdot L[\mathrm{cm}], \qquad x_{\parallel, \mathrm{max}} = x_{\perp, \mathrm{max}}^{2/3}.
\end{equation}

\subsection{$D_{\mu \mu}$ from fast modes}

In this Section, we instead consider the case of a cascade of fast-magnetosonic fluctuations. Analogously to the Alfv\'enic case, the details of the calculation leading to the associated pitch-angle scattering rate, $D_{\mu\mu}^{\rm F}$, are outlined.

\subsubsection{Normalization coefficient}
Again, to normalize the spectrum resulting from the simulations, we use Equation \eqref{eq:normalization_MHD_modes} for the corresponding spectrum of fast-magnetosonic turbulence obtained from the trace of the correlation tensor in \eqref{eq:fast_scalings} (we remind the reader that $\sum_{i=j} J_{ij} = \frac{k_i k_i + k_j k_j}{k_{\perp}^2} = 1$).
Since fast modes are found to be isotropic, we can rearrange the integral over the intertial range as $\int d^3 \bm{k} = \int_{L^{-1}}^{+ \infty} k^2 \, dk \int_{0}^{\pi} \sin{\alpha} \, d \alpha \int_{0}^{2 \pi} d \phi$.

The equation to solve to get the normalization is therefore:

\begin{equation*}
    C_a^{\rm F} \cdot 2\pi \int_{L^{-1}}^{+ \infty} k^2 \, dk \int_{0}^{\pi} \sin{\alpha} \, d \alpha \, k^{-7/2} = \frac{\langle \delta B^2 \rangle_{\mathrm{rms},L}}{B_0^2} = M_{\mathrm{A}}^2
\end{equation*}

From this, we get that $C_a^{\rm F} = \frac{M_{\mathrm{A}}^2 \, L^{-1/2}}{8 \pi}$ and finally:

\begin{equation}\label{eq:fast_spectrum_normalized}
    \mathcal{M}^{\rm F}_{ij} =  \frac{M_{\mathrm{A}}^2\, L^{1/2}}{8 \pi} J_{ij} k^{-7/2}.
\end{equation}

\subsubsection{Resonance function}\label{subsec:resonance_function}
The resonance function is the same presented in Equation \eqref{eq:resonant_function}, but split in two forms, as for scattering on fast modes contributions from both transit-time damping (TTD) and gyro-resonant interaction have to be taken into account. 

Gyroresonance corresponds to the case $n \neq 0$, and the resulting function is the same described for the Alfv\'en modes:
\[
    R_n (k_{\parallel} v_{\parallel} - \omega + n \Omega) = \frac{\sqrt{\pi}}{|k \xi| v_{\perp} M_{\mathrm{A}}^{1/2}} \cdot \exp \left(- \frac{(\mu + \frac{n}{x \xi R})^2}{(1 - \mu^2) M_{\mathrm{A}}} \right) \equiv \frac{\sqrt{\pi}}{|k \xi| v_{\perp} M_{\mathrm{A}}^{1/2}} \cdot E_n^{\rm G} \qquad (n \neq 0)
\]
where $\xi \equiv \cos{\alpha}$ is the ``pitch-angle'' of the wave vector associated to the turbulent fluctuations (\textit{i.e.}, $\alpha$ is the angle between $\bm{k}$ and $\bm{B}_0$).

Transit-time damping corresponds to $n=0$, in which case we can rearrange the argument of the exponential as $\frac{\cancel{k^2_{\parallel} v^2} \left( \mu - \frac{\omega}{k_{\parallel} v} \right)^2}{\cancel{k^2_{\parallel} v^2} (1 - \mu^2) M_{\mathrm{A}}} = \frac{\left( \mu - \frac{v_A}{\xi v} \right)^2}{(1 - \mu^2) M_{\mathrm{A}}} $, where the last step holds because the phase velocity of the fast waves is the same order of magnitude as the Alfv\'en speed, $\omega \approx k v_{\rm A}$, in the low-$\beta$ limit.

In this case, the resulting function is:
\[
    R_n (k_{\parallel} v_{\parallel} - \omega + n \Omega) = \frac{\sqrt{\pi}}{|k \xi| v_{\perp} M_{\mathrm{A}}^{1/2}} \cdot \exp \left(- \frac{(\mu - \frac{v_{\rm A}}{\xi v})^2}{(1 - \mu^2) M_{\mathrm{A}}} \right) \equiv \frac{\sqrt{\pi}}{|k \xi| v_{\perp} M_{\mathrm{A}}^{1/2}} \cdot E_n^{\rm T} \qquad (n=0).
\]

\subsubsection{Truncation scale}\label{subsec:truncation_scale}
The integral over the inertial range is truncated as soon as the fastest damping mechanism for the turbulent spectra comes into play. This eventually depends on the environment that we are considering. 

As discussed in \citet{Yan:2007uc}, in the warm ionized medium (WIM) ($|d| \lesssim 1 \, \mathrm{kpc}$) the gas is denser and colder with respect to the extended halo region ($d > 1 \, \mathrm{kpc}$). Therefore, in the WIM, besides the standard \textit{collisionless} damping, the \textit{collisional} damping is also present. Since viscous forces involve small-size eddies, only particles with small Larmor radii can experience them. This will eventually affect the low-energy range of the resulting spatial diffusion coefficient in the WIM.
In the extended halo region, on the other hand, only the collisionless damping is present, and this is why we expect $D(R)$ to be a monotonic function of $R$ in such environment.

To estimate the truncation scale in the two different environments, we look for the wave number at which the energy cascading rate of the turbulence equals the dissipation rate associated to that wave-number~\citep{Lazarian:2020cms}.

Following \citet{Yan:2007uc}, the collisionless truncation scale results:
\begin{equation}\label{eq:truncation_scale_collisionless}
    k_{\mathrm{max}} L = \frac{4 \, M_{\mathrm{A}}^4 \, \gamma \, \xi^2}{\pi \, \beta \, (1 - \xi^2)^2} \cdot \exp \left( \frac{2}{\beta \, \gamma \, \xi^2} \right),
\end{equation}
where $\gamma = \frac{m_p}{m_e}$ and $\beta = \frac{P_g}{P_B}$ is the ratio between the gas pressure and the magnetic pressure.

On the other hand, the collisional truncation scale is:
\begin{equation}\label{eq:truncation_scale_viscous}
    k_{\mathrm{max}} L = \begin{cases}
    x_c \, (1 - \xi^2)^{-2/3} & \quad \beta \ll 1\\
    x_c \, (1 - 3 \, \xi^2)^{-4/3} & \quad \beta \gg 1,
  \end{cases}
\end{equation}
where $x_c =  \left( \frac{6 \, \rho \, \delta V^2 \, L}{\eta_0 \, v_A} \right)^{2/3} \sim 10^6$ contains the ambient variables, with $\eta_0$ being a longitudinal viscosity~\citep{Yan:2007uc}. 

\subsubsection{Pitch-angle coefficient}
To calculate the contribution of the fast-magnetosonic modes to the pitch-angle diffusion coefficient, we plug in the spectrum \eqref{eq:fast_spectrum_normalized} in the following equation:
\begin{equation}\label{eq:pitch-angle_diffusion_fast_generic}
    D_{\mu \mu}^{\rm F} = \Omega^2 (1 - \mu^2) \int d^3 \bm{k} \sum^{+\infty}_{n=-\infty} \frac{\sqrt{\pi}}{|k_{\parallel}| v_{\perp} M_{\mathrm{A}}^{1/2}} \cdot E^{G,T}_n \left[ \frac{k_{\parallel}^2}{k^2} \, J'^{2}_{n}(z) \, I^{\rm F} (\bm{k}) \right],
\end{equation}
where now $z = k_{\perp} L R (1 - \mu^2)^{1/2} = k (1 - \xi^2)^{1/2} L R (1 - \mu^2)^{1/2} \equiv x R (1 - \xi^2)^{1/2} (1 - \mu^2)^{1/2}$.

With the usual notation $R \equiv v/(\Omega L) = (1-\mu^2)^{-1/2}r_{\rm L}/L$ and $k L \equiv x$, and using that $\xi^2/|\xi|$ is an even function, so that $\int_{-1}^{+1} d \xi \, \xi^2/|\xi| = 2 \int_{0}^{+1} d \xi \, \xi$, the general expression that computes the contributions from the fast modes to $D_{\mu \mu}$ is:
\begin{equation}\label{eq:pitch-angle_diffusion_fast}
    D_{\mu \mu}^{\rm F} = \frac{M_{\mathrm{A}}^{3/2} v \sqrt{\pi}}{2 R^2 L} (1 - \mu^2)^{1/2}  \int_{1}^{k_{\mathrm{max}} L(\xi)} dx \int_{0}^{+1} d \xi \, \xi \sum^{+\infty}_{n=-\infty} x^{-5/2} J'^{2}_{n}(z)\cdot E^{\rm G,T}_n
\end{equation}
where:
\[
E_n^{\rm G,T} =
    \begin{cases}
      E_n^{\rm T} = \exp \left(- \frac{(\mu - \frac{v_{\rm A}}{\xi v})^2}{(1 - \mu^2) M_{\mathrm{A}}} \right) & \quad (n=0)\\
      E_n^{\rm G} = \exp \left(- \frac{(\mu + \frac{n}{x \xi R})^2}{(1 - \mu^2) M_{\mathrm{A}}} \right) & \quad (n \neq 0).
    \end{cases}
\]

So, in the case of TTD interaction ($n=0$), we have:
\begin{equation}\label{eq:pitch-angle_diffusion_fast_TTD}
    D_{\mu \mu}^{{\rm F}, n=0} = \frac{M_{\mathrm{A}}^{3/2} v \sqrt{\pi}}{2 R^2 L} (1 - \mu^2)^{1/2}  \int_{1}^{k_{\mathrm{max}} L(\xi)} dx \int_{0}^{+1} d \xi \, \xi \, x^{-5/2} J^{2}_{1}(z) \cdot \exp \left(- \frac{(\mu - \frac{v_{\rm A}}{\xi v})^2}{(1 - \mu^2) M_{\mathrm{A}}} \right)
\end{equation}
where we used the property $J'_n(z) = \frac{1}{2} \left( J_{n-1}(z) - J_{n+1}(z) \right)$ to get $J'_0(z) = - J_1(z)$.

In the case of gyroresonant interaction ($n \neq 0$), we have:
\begin{equation}\label{eq:pitch-angle_diffusion_fast_Gyro}
    D_{\mu \mu}^{{\rm F}, n=1} + D_{\mu \mu}^{{\rm F}, n=-1} = \frac{M_{\mathrm{A}}^{3/2} v \sqrt{\pi}}{2 R^2 L} (1 - \mu^2)^{1/2}  \int_{1}^{k_{\mathrm{max}} L(\xi)} dx \int_{0}^{+1} d \xi \, \xi \, x^{-5/2} \left( \frac{J_0(z) - J_2(z)}{2} \right)^2 (E^{{\rm G},+} + E^{{\rm G},-}),
\end{equation}
where $E^{{\rm G},+} \equiv E^{{\rm G},+}_{n=1} = \exp \left(- \frac{(\mu + \frac{n}{x \xi R})^2}{(1 - \mu^2) M_{\mathrm{A}}} \right)$, $E^{G,-} \equiv E^{G,-}_{n=-1} = \exp \left(- \frac{(\mu - \frac{n}{x \xi R})^2}{(1 - \mu^2) M_{\mathrm{A}}} \right)$ and we used that $J'^{2}_n(z) = J'^{2}_{-n}(z)$.

\subsection{$D_{\mu\mu}$ from slow modes}

For completeness, we also report the calculations of the pitch-angle coefficient of the magnetosonic slow modes, namely the following expression:
\begin{equation}\label{eq:pitch-angle_diffusion_slow_generic}
    D_{\mu \mu}^{\rm S} = \Omega^2 (1 - \mu^2) \int d^3 \bm{k} \sum^{+\infty}_{n=-\infty} \frac{\sqrt{\pi}}{|k_{\parallel}| v_{\perp} M_{\mathrm{A}}^{1/2}} \cdot E^{G,T}_n \left[ \frac{k_{\parallel}^2}{k^2} \, J'^{2}_{n}(z) \, I^{\rm S} (\bm{k}) \right],
\end{equation}
where we want to adopt the same notation used for the Alfvén modes, separating the parallel and perpendicular wave-number components, as with respect to the regular magnetic field, $z = k_{\perp} L R (1 - \mu^2)^{1/2} \equiv x_{\perp} R (1 - \mu^2)^{1/2}$.

The statistics of the slow modes is similar to that of the Alfvén modes, as indicated in Equations \eqref{eq:alfven_scaling_sub}-\eqref{eq:alfven_scaling_super}, while, on the other hand, they can interact with cosmic-ray particles by means of both TTD and gyro-resonance. Therefore their treatment involves parts of the calculations already detailed for the other two MHD modes. In particular:
\begin{itemize}
    \item the normalized correlation tensors $\mathcal{M}^{S}_{ij}$ for both the sub-alfvénic and super-alfvénic cases are the same calculated for the Alfvèn modes, reported in Equations \eqref{eq:alfven_spectrum_normalized} and \eqref{eq:alfven_spectrum_normalized_super}, respectively;
    \item the resonance function is the same as for the fast modes, discussed in Section \ref{subsec:resonance_function}, conveniently rewritten as follows to account for the present notation:
    \begin{equation*}
        R_n^{\rm G,T} =
    \begin{cases}
      R_n^{\rm T} \equiv \frac{\sqrt{\pi}}{|k_{\parallel}| v_{\perp} M_A^{1/2}} \cdot E_n^T = \frac{\sqrt{\pi}}{|k_{\parallel}| v_{\perp} M_A^{1/2}} \cdot \exp \left(- \frac{( \mu - \frac{\omega}{k_{\parallel} v} )^2}{(1 - \mu^2) M_{\mathrm{A}}} \right) & \quad (n=0)\\[6pt]
      R_n^{\rm G} \equiv \frac{\sqrt{\pi}}{|k_{\parallel}| v_{\perp} M_A^{1/2}} \cdot E_n^G = \frac{\sqrt{\pi}}{|k_{\parallel}| v_{\perp} M_A^{1/2}} \cdot \exp \left(- \frac{(\mu + \frac{n}{k_{\parallel} R L})^2}{(1 - \mu^2) M_{\mathrm{A}}} \right) & \quad (n \neq 0);
    \end{cases}
    \end{equation*}
    \item the truncation scale is also the same as that discussed for the fast modes, in Section \ref{subsec:truncation_scale}.
\end{itemize}

\subsubsection{Pitch-angle coefficient}\label{subsubsec:pitch-angle_slow}
To calculate the $D_{\mu \mu}$ caused by the slow modes, we account for the sub- and super-alfvénic nature of the injected cascade, separately.

\vspace{0.3cm}
\hspace{-0.6cm}\textit{\textbf{Sub-alfv\'enic case: $\bm{M_{\mathrm{A}} \leq 1}$}}. The general expression that calculates the contribution from the slow modes to the pitch-angle coefficient is then written as follows:
\begin{equation}\label{eq:slow_modes_general_subalfvenic}
    D_{\mu \mu}^{\rm S} = \frac{2 \sqrt{\pi} v \sqrt{1 - \mu^2\,} \, M_{\rm A}^{5/6}}{3 R^2 L} \int_{\mathbb{R}^{+}} d x_{\perp} \int d x_{\parallel} \sum^{+ \infty}_{n=- \infty} \frac{x_{\perp}^{-7/3}}{\left( x^2_{\parallel} + x^2_{\perp} \right)} \cdot \exp \left(- \frac{x_{\parallel}}{M_{\rm A}^{4/3} \, x_{\perp}^{2/3}} \right) E^{G,T}_n,
\end{equation}
where we used the even parity of the integrating function to restrict only to the positive axis and the integral boundaries are the ones discussed in Section \ref{subsubsec:pitch-angle_alfven}.

In the case of TTD interaction ($n=0$), the expression above reads:
\begin{equation}
    D_{\mu \mu}^{\mathrm{S}, \mathrm{sub}, n=0} = \frac{2 \sqrt{\pi} v \sqrt{1 - \mu^2\,} \, M_{\rm A}^{5/6}}{3 R^2 L} \int_{\mathbb{R}^{+}} d x_{\perp} \int d x_{\parallel} \frac{x_{\perp}^{-7/3}}{\left( x^2_{\parallel} + x^2_{\perp} \right)} J_1^2(z) \cdot \exp \left(- \frac{x_{\parallel}}{M_{\rm A}^{4/3} \, x_{\perp}^{2/3}} - \frac{( \mu - \frac{\omega}{x_{\parallel} R \Omega} )^2}{(1 - \mu^2) M_{\mathrm{A}}} \right).
\end{equation}

In the case of gyro-resonant scattering ($n \neq 0$), on the other hand, Equation \eqref{eq:slow_modes_general_subalfvenic} is written as follows:
\begin{equation}
    D_{\mu \mu}^{\mathrm{S}, \mathrm{sub}, n=1} + D_{\mu \mu}^{\mathrm{S}, \mathrm{sub}, n=-1} = \frac{2 \sqrt{\pi} v \sqrt{1 - \mu^2\,} \, M_{\rm A}^{5/6}}{3 R^2 L} \int_{\mathbb{R}^{+}} d x_{\perp} \int d x_{\parallel} \frac{x_{\perp}^{-7/3}}{\left( x^2_{\parallel} + x^2_{\perp} \right)} \left( \frac{J_0(z) - J_2(z)}{2} \right)^2 \cdot \exp \left(- \frac{x_{\parallel}}{M_{\rm A}^{4/3} \, x_{\perp}^{2/3}} \right) \cdot \left( E^{G, +} + E^{G, -} \right)
\end{equation}
with obvious meaning of the terms $E^{G,+}$ and $E^{G,-}$.

\vspace{0.3cm}
\hspace{-0.6cm}\textit{\textbf{Super-alfv\'enic case: $\bm{M_{\mathrm{A}} > 1}$}}. In the case of super-alfvénic turbulence injected, the general expression for the pitch-angle coefficient is the following:
\begin{equation}\label{eq:slow_modes_general_superalfvenic}
    D_{\mu \mu}^{\rm S} = \frac{2 \sqrt{\pi} v \sqrt{1 - \mu^2\,} \, M_{\rm A}^{1/2}}{3 R^2 L} \int_{\mathbb{R}^{+}} d x_{\perp} \int d x_{\parallel} \sum^{+ \infty}_{n=- \infty} \frac{x_{\perp}^{-7/3}}{\left( x^2_{\parallel} + x^2_{\perp} \right)} \cdot \exp \left(- \frac{x_{\parallel}}{M_{\rm A} \, x_{\perp}^{2/3}} \right) E^{G,T}_n.
\end{equation}

In the case of TTD particle-wave interaction ($n=0$), this becomes:
\begin{equation}
    D_{\mu \mu}^{\mathrm{S}, \mathrm{sub}, n=0} = \frac{2 \sqrt{\pi} v \sqrt{1 - \mu^2\,} \, M_{\rm A}^{1/2}}{3 R^2 L} \int_{\mathbb{R}^{+}} d x_{\perp} \int d x_{\parallel} \frac{x_{\perp}^{-7/3}}{\left( x^2_{\parallel} + x^2_{\perp} \right)} J_1^2(z) \cdot \exp \left(- \frac{x_{\parallel}}{M_{\rm A} \, x_{\perp}^{2/3}} - \frac{( \mu - \frac{\omega}{x_{\parallel} R \Omega} )^2}{(1 - \mu^2) M_{\mathrm{A}}} \right).
\end{equation}

In the case of gyro-resonant interaction ($n \neq 0$), instead, Equation \eqref{eq:slow_modes_general_superalfvenic} becomes:
\begin{equation}
    D_{\mu \mu}^{\mathrm{S}, \mathrm{sub}, n=1} + D_{\mu \mu}^{\mathrm{S}, \mathrm{sub}, n=-1} = \frac{2 \sqrt{\pi} v \sqrt{1 - \mu^2\,} \, M_{\rm A}^{1/2}}{3 R^2 L} \int_{\mathbb{R}^{+}} d x_{\perp} \int d x_{\parallel} \frac{x_{\perp}^{-7/3}}{\left( x^2_{\parallel} + x^2_{\perp} \right)} \left( \frac{J_0(z) - J_2(z)}{2} \right)^2 \cdot \exp \left(- \frac{x_{\parallel}}{M_{\rm A} \, x_{\perp}^{2/3}} \right) \cdot \left( E^{G, +} + E^{G, -} \right).
\end{equation}

\bibliography{sample63}{}
\bibliographystyle{mnras}

\end{document}